\documentclass[referee,sn-nature]{sn-jnl}

\usepackage[export]{adjustbox}
\usepackage{graphicx}%
\usepackage{multirow}%
\usepackage{amsmath,amssymb,amsfonts}%
\usepackage{amsthm}%
\usepackage{mathrsfs}%
\usepackage[title]{appendix}%
\usepackage{xcolor}%
\usepackage{textcomp}%
\usepackage{manyfoot}%
\usepackage{xurl}%
\usepackage{booktabs}%
\usepackage{newunicodechar}
\newunicodechar{−}{-}
\usepackage{algorithm}%
\usepackage{algorithmicx}%
\usepackage{algpseudocode}%
\usepackage{listings}
\usepackage{comment}
\usepackage{overpic}
\usepackage{subcaption}
\usepackage{xr-hyper}
\usepackage{hyperref}
\usepackage{changepage}
\newsavebox{\panelbox}
\newcommand{\panel}[2]{%
  \sbox{\panelbox}{\includegraphics[width=\linewidth]{#2}}%
  \makebox[0pt][r]{%
    \raisebox{\dimexpr\ht\panelbox-\ht\strutbox\relax}[0pt][0pt]{\textbf{(#1)}}%
    \hspace{0.35em}}%
  \usebox{\panelbox}%
}

\theoremstyle{thmstyleone}%
\theoremstyle{thmstyletwo}%

\theoremstyle{thmstylethree}%

\newcommand{\CR}[1]{}
\newcommand{\IC}[1]{}
\newcommand{\PH}[1]{}

\begin{document}

\title[Article Title]{Predicting Mutational Signature Exposures from H\&E Whole Slide Images: A Pan-Cancer Feasibility Study}


\author[1]{\fnm{Flavio} \sur{Sartori}}
\email{flavio.sartori@unito.it}

\author[2]{\fnm{Cesare} \sur{Rollo}}
\email{cesare.rollo@di.ku.dk}

\author[1]{\fnm{Isabella} \sur{Caranzano}}
\email{isabella.caranzano@unito.it}

\author[1]{\fnm{Tiziana} \sur{Sanavia}}
\email{tiziana.sanavia@unito.it}

\author[1]{\fnm{Piero} \sur{Fariselli}}
\email{piero.fariselli@unito.it}
\equalcont{These authors jointly supervised this work.}

\author[3]{\fnm{Corrado} \sur{Pancotti}}
\email{corrado.pancotti@helmholtz-munich.de}
\equalcont{These authors jointly supervised this work.}

\affil[1]{\orgdiv{University of Torino}, \orgname{Medical Sciences}, \orgaddress{\street{Via Santena 19}, \city{Torino}, \postcode{10123}, \state{Piemonte}, \country{Italy}}}

\affil[2]{\orgdiv{Center for Health Data Science}, \orgname{Department of Computer Science}, \city{University of Copenhagen}, \country{Denmark}}

\affil[3]{\orgdiv{Helmholtz Munich}, \orgname{Helmholtz AI}, \orgaddress{\street{Ingolstädter Landstraße 1}, \city{Munich}, \postcode{85764}, \country{Germany}}}

\abstract{Mutational signatures reveal cancer-driving processes with clinical relevance: MMR-deficient tumors respond better to immunotherapy, HRD tumors are sensitive to PARP inhibitors, and POLE-mutant tumors often have high mutation burdens influencing treatment response. Yet signature profiling remains limited by the cost, complexity, and turnaround time of whole-genome sequencing (WGS) or whole-exome sequencing. Histopathology is widely available and cost-effective, and H\&E morphology has been linked to MSI, HRD, POLE-related processes, and driver mutations. Whether histology can recover the broader landscape of mutational signature exposures in a pan-cancer setting remains unknown. Here we introduce Hist2Sig, a deep learning framework predicting exposures to 30 COSMIC SBS signatures from H\&E-stained slides. Trained on matched WGS and histology data from 7,063 TCGA patients across 29 cancer types, Hist2Sig was compared with a tumor type-only baseline to separate morphology-derived signal from tissue-of-origin priors. In internal cross-validation, it recovered relative signature compositions across most tumor types and achieved higher mean top-three overlap than the baseline in 19 of 29, with the largest gains in COAD, PRAD, GBM, and UCEC. It also recurrently recovered signatures of poorly characterized etiology, including SBS8, SBS12, SBS39, and SBS40a. In an external CPTAC cohort of 193 patients across five tumor types, Hist2Sig retained moderate concordance with observed compositions and outperformed the baseline in glioblastoma and pancreatic adenocarcinoma. These results support the feasibility of inferring mutational signature exposures from routine histology, while highlighting variation across tumor types. Hist2Sig provides a foundation for tumor-specific models that could complement sequencing in pre-sequencing triage or support decisions where WGS is unavailable.}


\maketitle

\section{Introduction}

Cancer genomes accumulate mutations throughout tumor development, each
reflecting distinct biological processes and environmental exposures. These
mutations group into characteristic patterns called ``mutational signatures,''
which serve as biological fingerprints revealing the underlying mechanisms
of tumorigenesis. More than 80 unique mutational signatures have been
identified to date \cite{alexandrov2013signatures,alexandrov2020repertoire},
linking mutations to various factors such as ultraviolet radiation, smoking,
and deficiencies in DNA repair pathways including homologous recombination
and mismatch repair \cite{boysen2025investigating,koh2021mutational,hwang2025comprehensive,alexandrov2015mutational}.
Many of these signatures hold significant clinical implications: tumors
displaying the homologous recombination deficiency (HRD) signature respond
well to PARP inhibitors and platinum-based chemotherapy, while mismatch
repair deficiency (MMRd) signatures predict favorable responses to
immunotherapy \cite{brady2022therapeutic,ma2018therapeutic,alexandrov2016mutational,secrier2016mutational}.
Despite this clinical value, and despite the availability of established
computational tools for signature extraction such as \cite{blokzijl2018mutationalpatterns,gehring2015somaticsignatures,pancotti2024muse,jin2024accurate,islam2022uncovering,diaz2018mutational,pancotti2023unravelling}, routine identification of
mutational signatures in clinical practice remains limited. The
bottleneck is not algorithmic but upstream: whole-genome sequencing
(WGS) and whole-exome sequencing (WES), which are required to generate
the mutation catalogs these tools operate on, are still constrained by
cost, technical complexity, and turnaround time. As a result, these
methods remain largely confined to research environments or specialized
clinical settings, slowing the widespread adoption of signature-driven
precision oncology.

Conversely, histopathological examination of tissue slides is standard
practice in clinical oncology and is rapid, cost-effective, and universally
accessible. Advances in artificial intelligence and digital pathology have
recently demonstrated that histology images, beyond their conventional
diagnostic role, contain morphological patterns that reflect the underlying
genetic alterations \cite{cifci2022artificial,unger2024deep,krause2021deep}.
Deep learning models can successfully predict critical genomic traits
such as microsatellite instability (MSI), homologous recombination deficiency, and even specific driver
mutations directly from routinely stained histological images
\cite{bergstrom2024deep,wang2024deep,pizurica2024digital,liu2025mmrnet}.

Recent work has advanced this direction significantly.
In \cite{bergstrom2024deep} the authors trained a deep learning model to predict HRD
status and platinum response from histology, while others have addressed
MSI prediction in endometrial cancer \cite{wang2024deep,liu2025mmrnet} and
tissue-specific gene expression inference from H\&E slides
\cite{pizurica2024digital}. These efforts share a common design: a single
binary or continuous endpoint, typically in one or two tissues, with a
model architecture tailored to that endpoint. Whether the same
morphological information can support a joint, pan-cancer prediction of
the full landscape of mutational signatures, including the substantial
fraction of catalogued signatures with weakly characterized or unknown
etiology \cite{alexandrov2013signatures,alexandrov2020repertoire}, is
methodologically and biologically a distinct question, and the one we
address here.

A central premise of this work is that mutational processes leave traces
detectable at the tissue level. For a minority of mutational processes, this link
is well established. MMRd tumors exhibit characteristic
tumor-infiltrating lymphocytes and medullary features
\cite{greenson2009pathologic}, HRD-associated breast
cancers show distinctive growth patterns including laminated fibrosis
and high tumor cell density \cite{lazard2022deep, bergstrom2024deep},
and UV-induced damage co-occurs with recognizable solar elastosis in
the surrounding tissue \cite{thomas2010associations}. For most
signatures, however, no such correlate has been documented, and for the
substantial fraction of catalogued signatures with unknown etiology the
question is still open. A pan-cancer model offers a natural empirical
setting in which to probe whether such correlates exist at all, and to
do so jointly across signatures rather than one endpoint at a time.

A central challenge in this setting is that mutational processes and
tissue of origin are strongly entangled. Several dominant signatures are
nearly deterministic given the tumor type. SBS4 in lung adenocarcinoma
is linked to tobacco exposure
\cite{alexandrov2016mutational}, SBS7a/b dominate cutaneous melanoma due to
UV damage, and clock-like signatures such as SBS1 and SBS5 are ubiquitous
across tissues with relatively stereotyped contributions per tumor type
\cite{alexandrov2020repertoire}. A pan-cancer histology-to-signature model
could therefore appear to predict mutational processes while in fact
recovering little more than tissue identity. Decoupling the two requires
benchmarking any such model against a tumor type–only baseline that captures exactly this prior, a comparison that, to our knowledge, has not been systematically reported in prior works.

In this study, we introduce Hist2Sig, a deep learning framework trained on
whole genome sequenced tumors from The Cancer Genome Atlas (TCGA) across 29
cancer types, enabling a pan-cancer analysis of this task. Hist2Sig predicts exposure profiles across 30 selected COSMIC SBS mutational signatures directly from histopathological images, and we evaluate its ability to recover the dominant signatures contributing the largest fraction of mutations within each tumor. We benchmark the model against a
tumor type--only Random Forest baseline to isolate the contribution of
morphology beyond tissue-of-origin priors, and we externally validate on
an independent cohort from the Clinical Proteomic Tumor Analysis
Consortium (CPTAC), processed under different scanning and staining
protocols. In several tumor types Hist2Sig outperforms the
baseline by substantial margins, and recurrently recovers signatures of
incompletely characterized etiology and without a morphological association, most notably SBS8, SBS12, SBS18, SBS39 and SBS40a,
which the baseline cannot identify. We further show that the model
captures meaningful sample level variation within tumor types for a subset
of clinically relevant signatures.

\begin{figure*}[t]              
    \includegraphics[width=1.\linewidth]{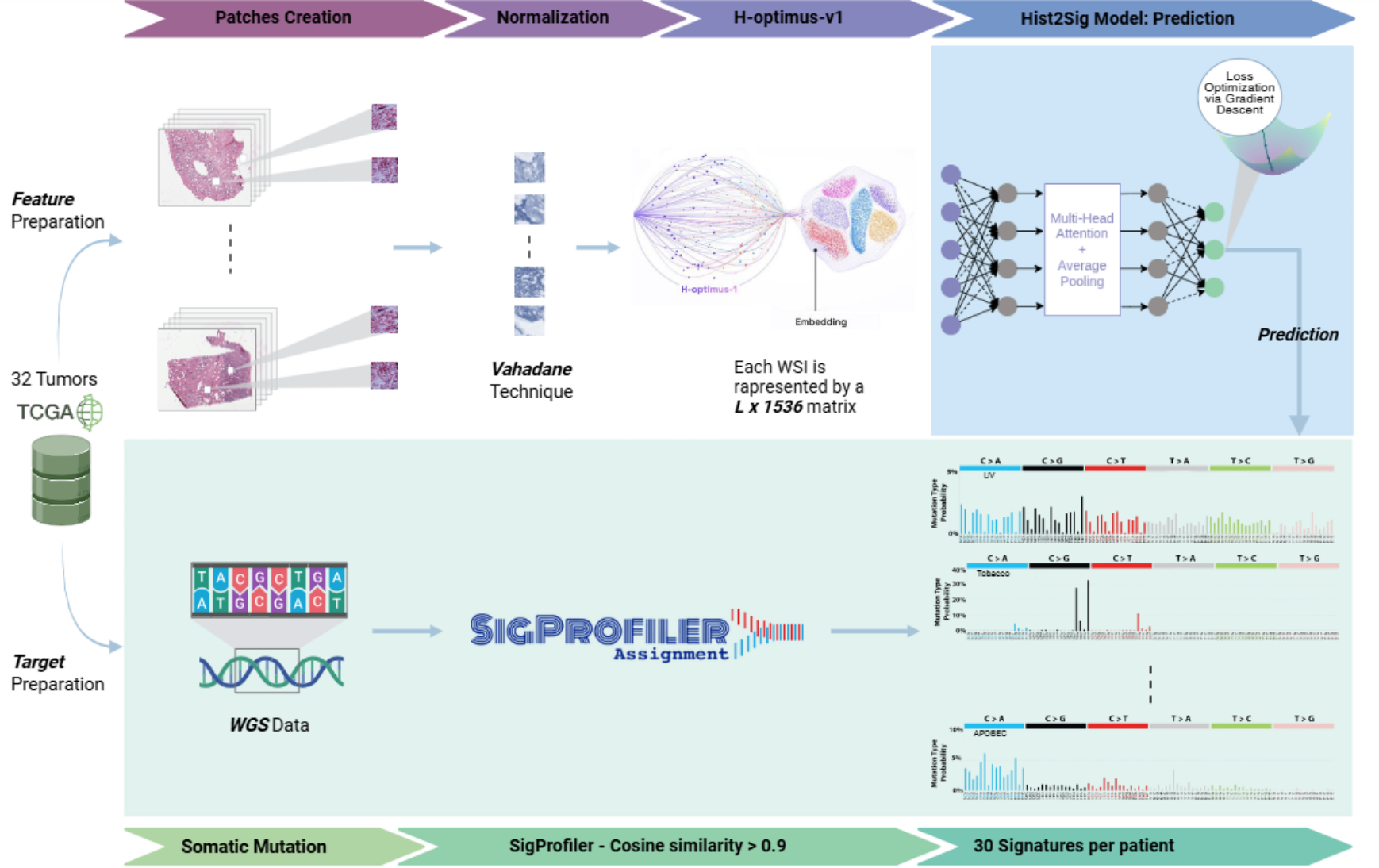}
    \centering
    \caption{\textbf{Overview of the Hist2Sig framework}.(Top) Feature preparation pipeline: diagnostic whole-slide images (WSIs) from 32 TCGA tumor types are tiled, stain-normalized using the Vahadane method, and encoded with H-optimus-v1, a 1.1B-parameter Vision Transformer, producing a patch-level feature matrix of size N × 1536 per slide. (Bottom) Target preparation pipeline: somatic single-nucleotide variants from matched WGS data are decomposed into exposures to 30 COSMIC SBS mutational signatures using SigProfilerAssignment, retaining only samples with reconstruction cosine similarity $\geq 0.9$. The Hist2Sig model takes the WSI features as input and predicts the corresponding signature exposure vector via an attention-based Multiple Instance Learning architecture}
    \vspace{5pt}  
    \label{fig:hist2sig}
\end{figure*}

We frame this work as a feasibility study. The goal is to establish whether the task is tractable in a
pan-cancer setting and to identify the tumor types and signatures where
morphology carries information beyond tissue of origin, providing a
foundation for the development of dedicated tumor type--specific models
that may achieve improved performance. More broadly, an approach of this
kind could complement rather than replace sequencing, for example as a
pre-sequencing triage tool to prioritize patients most likely to benefit
from genomic profiling, or to extend signature-informed decision-making to
settings where WGS is not routinely available.
An overview of the Hist2Sig framework is shown in Figure \ref{fig:hist2sig}.

\newpage
\section{Results}

\subsection{Mutational signature analysis}\label{mutsig}

We ran SigProfilerAssignment on both cohorts on each vcf file thus assigning COSMIC v3.4 mutational signatures to each sample individually. Then we performed the preprocessing described in the Methods (Section~\ref{subsec21}). After filtering and H\&E WSI matching, the final TCGA cohort comprised 7,063 patients across 29 tumor types and 30 retained mutational signatures, including 20 of known etiology and 10 of unknown origin; the final CPTAC evaluation cohort comprised 193 samples across five tumor types.
In Supplementary Figure S1  and S2 we show the distribution of the 30 signatures extracted by SigProfilerAssignment in the filtered TCGA cohort and CPTAC respectively together with the distribution of tumor types.

\subsection{Hist2Sig Prediction of Mutational Signature Exposure on TCGA from H\&E}\label{subsec:tcga_pred}

\begin{figure}[h!]    \includegraphics[width=1.\linewidth]{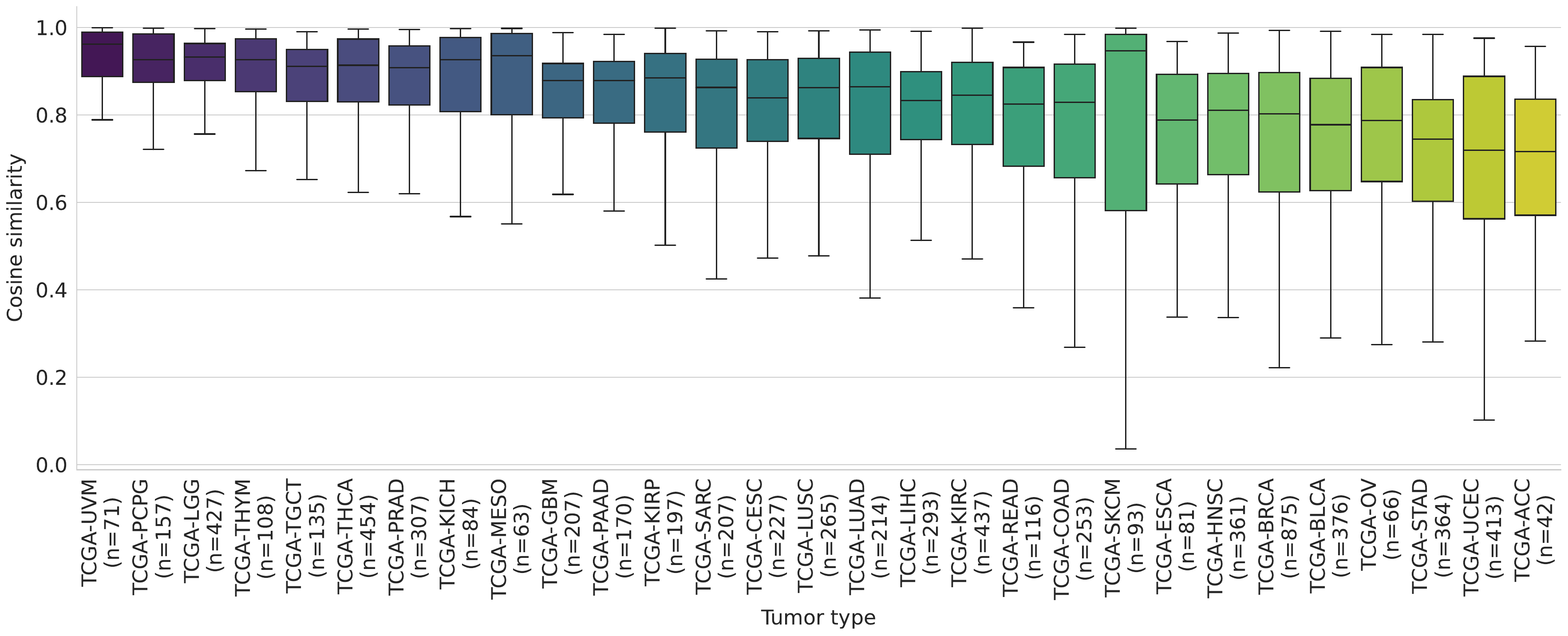}
    \caption{\textbf{Distribution of cosine similarities in the TCGA cohort.} Cosine similarity between predicted and ground-truth mutational signature exposure profiles, by tumor type. Box plots show median and interquartile range.}
    \label{fig:cosine_TCGA}
\end{figure}

To quantitatively assess Hist2Sig performance, we first computed for each tumor type the distribution of cosine similarity between predicted and ground-truth exposure profiles, each sample being represented as a 30-dimensional signature exposure vector (Section~\ref{sig_eval}). As shown in Figure~\ref{fig:cosine_TCGA}, most tumor types reach high values, with many medians above 0.8, indicating that the model can accurately recover the relative proportions of mutational signatures; however greater variability is observed for UCEC, BRCA and SKCM, likely reflecting the higher heterogeneity of their mutational landscapes.

\begin{figure}[t]
\centering
\includegraphics[width=1\linewidth]{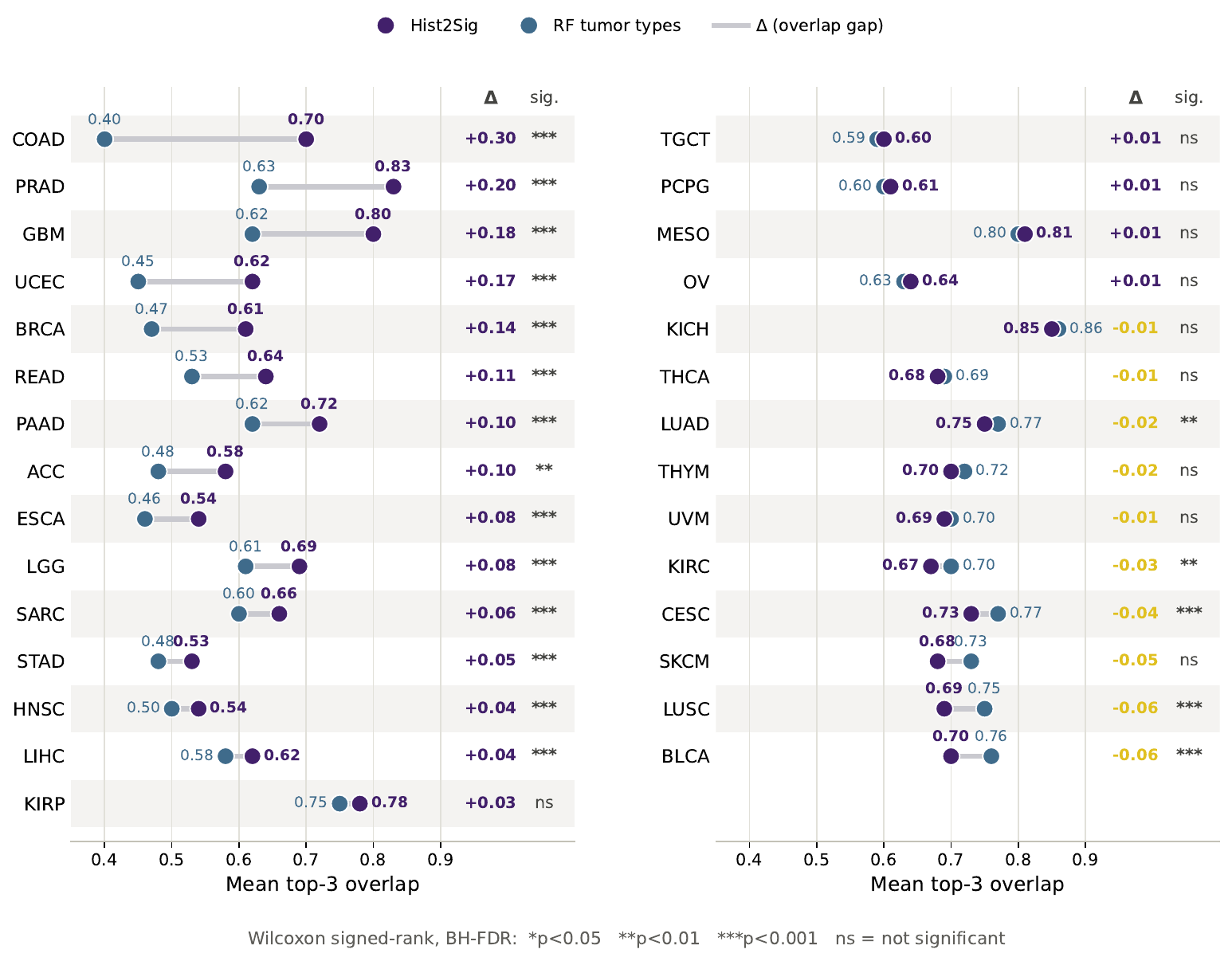}
\caption{\textbf{Mean top-3 overlap score for Hist2Sig and the Random
    Forest baseline across tumor types in the TCGA cohort.} For each
    tumor type, the two dots show the mean fraction of the three
    dominant observed signatures correctly identified among the top-3
    predicted, for Hist2Sig (violet) and a Random Forest trained on
    tumor type labels (blue); the grey connector spans the gap between
    them. The $\Delta$ column reports the pairwise difference in mean
    overlap (Hist2Sig $-$ RF), coloured purple where Hist2Sig
    outperforms the baseline and red where the baseline prevails.
    Statistical significance of the difference between the two models is
    indicated in the adjacent column (Wilcoxon signed-rank test,
    Benjamini--Hochberg FDR correction; $\ast p < 0.05$,
    $\ast\ast p < 0.01$, $\ast\ast\ast p < 0.001$; ns, not significant).
    Tumor types are ordered by the difference in mean overlap between
    the two models, split across the two panels (left: largest Hist2Sig
    advantage; continuing to the right).}
\label{fig:mean_top_TCGA} 
\end{figure}

We then evaluate the model’s ability to recover dominant mutational signatures, as identifying the processes responsible for most mutations is often more clinically relevant than predicting every mutational process. Figure \ref{fig:mean_top_TCGA} presents the mean top-three overlap score for each tumor type, comparing Hist2Sig with a Random Forest baseline trained using tumor-type labels to predict the three signatures with the highest activity.
Tumor types are ordered by the difference in performance between the two models. Hist2Sig achieves a higher mean overlap score in 19 of the 29 tumor types, with the largest improvements observed in COAD ($+0.30$), PRAD ($+0.20$), GBM ($+0.18$), and UCEC ($+0.17$). In the 10 tumor types for which the Random Forest baseline performs better, the differences are consistently smaller, at most $0.06$, and few are statistically significant. This asymmetry suggests that Hist2Sig provides substantial gains when it outperforms the baseline, whereas tumor-type information alone confers only modest advantages in the remaining cases.

The complete per sample overlap score distributions for each tumor type are reported in Supplementary Figure S3. 
Here, it is possible to notice that for the tumor types where the performance gap is most pronounced, the Random Forest baseline frequently fails to correctly recover all three dominant signatures, as it assigns identical predictions to every sample within the same tumor type and thus cannot adapt to sample level variations in the mutational landscape.

\begin{figure}[p]
  \centering
\includegraphics[width=0.9\linewidth]{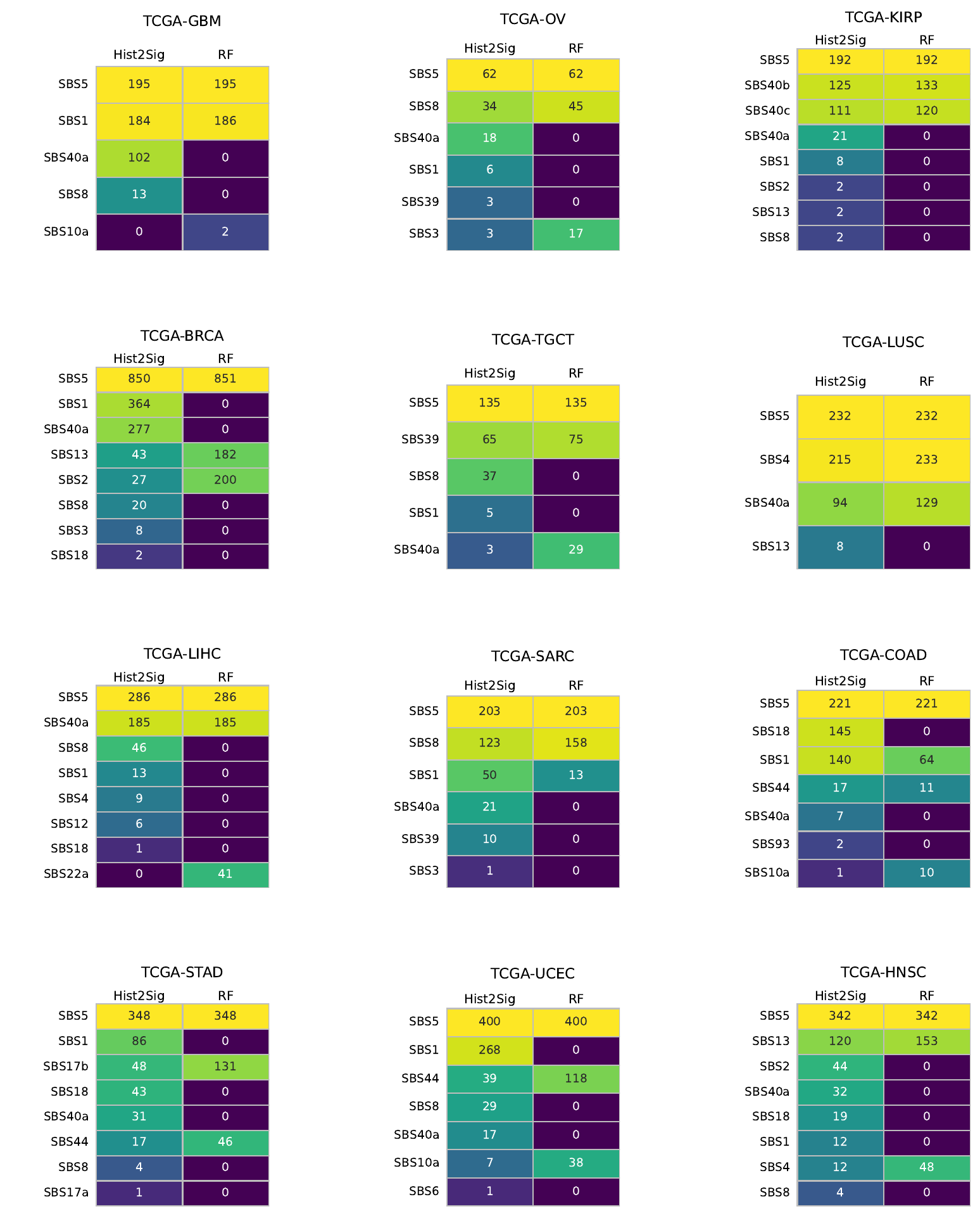}
  \caption{\textbf{Signature diversity in the top 3 dominant signatures for selected tumor types in the TCGA cohort.} For each tumor type and signature, cell values indicate the number of samples in which each signature was correctly identified among the top three by Hist2Sig and by the Random Forest baseline (RF); color intensity (log scale) reflects the magnitude of the count. Only the 12 tumor types with the greatest heterogeneity in predicted signatures are shown; complete results for all tumor types are reported in Supplementary Figure S4.}
  \label{fig:diversity_tcga}
\end{figure}

In addition, to assess whether Hist2Sig captures a more diverse mutational landscape we examined which signatures each model correctly identifies among the top three, across samples and its diversity. Figure~\ref{fig:diversity_tcga} shows that SBS1, SBS8, SBS18, SBS39 and SBS40a are recurrently recovered as dominant signatures by Hist2Sig while the Random Forest baseline missed them in most tumor types. SBS1 and SBS40a are clock-like signatures associated with aging and cumulative cell division history, processes with plausible histomorphological correlates in tissue turnover, proliferative state and architectural remodeling \cite{alexandrov2015clock}; SBS18 is associated with oxidative DNA damage through 8-oxoguanine lesions \cite{kucab2019compendium}, and SBS8, of only partially characterized etiology, has been linked to oxidative stress and defective DNA repair mechanism \cite{singh2020mutational}.

Hist2Sig also recovers dominant signatures that the baseline misses entirely: SBS2, associated with APOBEC mutagenesis, in 44 HNSC samples; SBS3, linked to HRD, in 8 BRCA samples; SBS12, a liver-specific signature, in LIHC; and SBS39 across sarcoma, ovarian and neuroendocrine tumors. The etiology of SBS39 is classified as unknown in COSMIC, although a recent preprint suggests a possible association with HRD related process \cite{ding2024cosmic}.
These results suggest that the morphological features extracted from whole slide images provide meaningful information about the dominant mutational processes beyond what can be inferred from tumor type alone.

\begin{figure}[h!]
\centering \includegraphics[width=1.\linewidth]{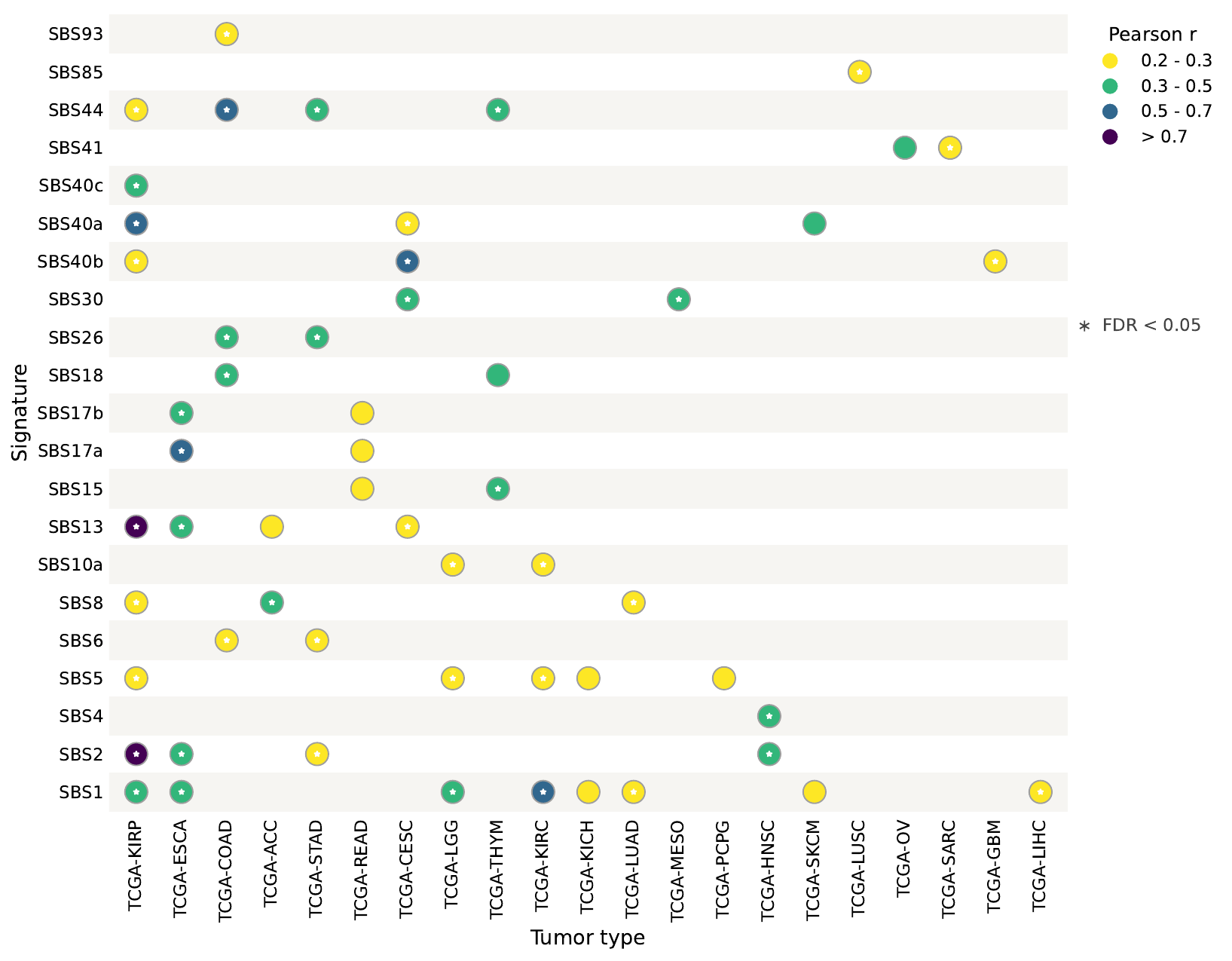}
    \caption{\textbf{Per signature Pearson correlation between predicted and ground-truth mutational signature exposures across tumor types in the TCGA cohort}. Each bubble represents the Pearson correlation coefficient between predicted and ground-truth exposures for a given signature within a specific tumor type, computed across all samples of that tumor type. Only correlations with $r \ge 0.2$ are shown. Bubble color indicates the magnitude of the correlation according to four bins. Asterisks denote statistically significant correlations after Benjamini-Hochberg false discovery rate correction (FDR $<$ 0.05). Tumor types are ordered by the number of signatures with $r \ge 0.2$.}
    \vspace{15pt}  
    \label{fig:bubble_tcga}
\end{figure}

Further evidence of Hist2Sig’s ability to capture morphological information comes from its recovery of meaningful intra-tumor sample-level variation in mutational processes. Figure \ref{fig:bubble_tcga} shows the distribution of Pearson correlations between predicted and ground-truth exposures across samples and signatures within each tumor type, restricted to correlations with $r \ge 0.2$. Bubbles marked with an asterisk indicate statistically significant correlations after FDR correction ($p < 0.05$). Overall, Hist2Sig captures meaningful sample-level variation in several clinically relevant signatures across multiple tumor types. In particular, SBS2 and SBS13, both associated with APOBEC mutagenesis, show the highest correlations ($r > 0.7$) in KIRP, with significant correlations also observed in ESCA, COAD, and CESC. The clock-like signature SBS1 achieves significant correlations across a broad range of tumor types, including ESCA, KIRC, KIRP, LGG and LUAD, suggesting that the model consistently captures age and proliferation related morphological features. SBS18, linked to oxidative DNA damage, shows widespread moderate correlations across several tumor types, while SBS17a and SBS17b, associated with gastric and esophageal cancers, exhibit notable correlations in STAD and READ. SBS4, attributed to tobacco smoking, reaches a significant correlation in HNSC, and SBS44, related to defective mismatch repair, is significantly correlated in multiple tumor types including COAD, READ, and KIRC. Overall, the bubble plot demonstrates that Hist2Sig captures intra tumor type heterogeneity at the level of individual signatures. 

\subsection{Morphological specificity of signature attribution maps}\label{sec:morphspec}

The analyses presented so far demonstrate that Hist2Sig recovers signature exposures beyond the information provided by tumor type alone; however, they do not reveal where the supporting evidence is located within the slide. 
We therefore investigated whether the attribution map for a given signature is spatially distinct from those of other signatures co-occurring within the same slide, and whether this distinctness becomes more pronounced when the corresponding mutational process is genomically active. Patches were ranked by the attention-weighted gradient attribution score $c_i^{(k)}$,
which combines the attention assigned to a patch with the gradient
of the predicted exposure for signature $k$ with respect to that patch embedding
(Section~\ref{subsec:morph_methods}). For each tumor type and each signature recurrently detected within it, we compared patients for whom the signature ranked among the three dominant processes with those for whom it did not. Both groups were restricted to correctly classified samples. Spatial specificity was quantified using the Jaccard overlap, $J_K$, and the uniqueness, $U_K$, of the top-$K$ attributed patches.

\begin{figure}[h!]
\centering{  \includegraphics[width=1.\linewidth]{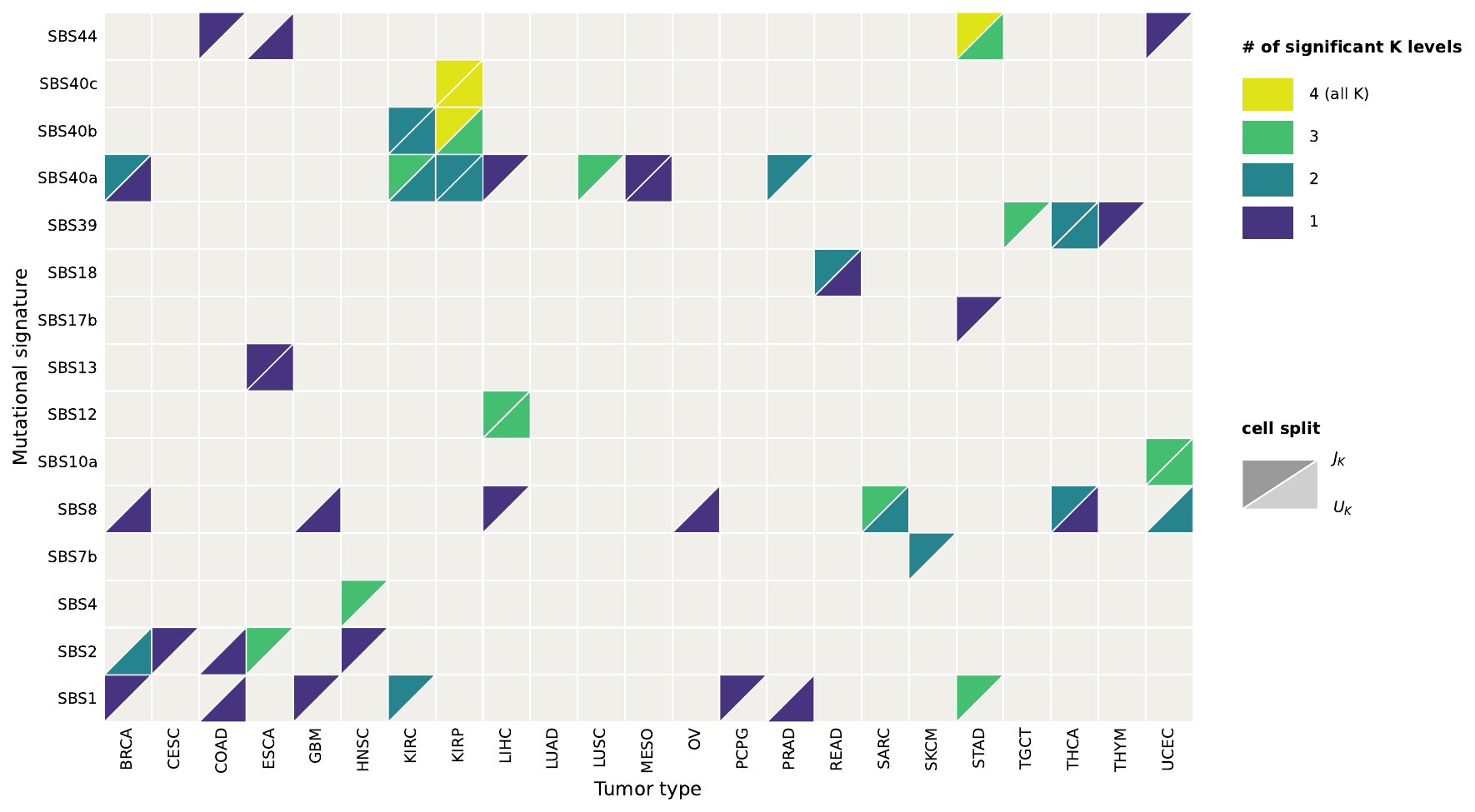}}
\caption{\textbf{Spatial specificity of signature attribution maps
    across tumor types in the TCGA cohort.} For each tumor
    type--signature pair, true-positive and true-negative samples were
    compared using two statistics computed over the top-$K$ attributed
    patches: the Jaccard overlap $J_K$ with patches attributed to
    co-dominant signatures in the same sample (upper-left triangle of
    each cell) and patch uniqueness $U_K$ (lower-right triangle). Lower
    $J_K$ and higher $U_K$ indicate greater signature-specific
    localization. A cell (or triangle) is shown only where the positive
    group was significantly more specific than the negative group
    (one-sided Mann--Whitney $U$ test, $p < 0.05$); all significant
    differences favored the positive group. Color encodes the number of
    top-$K$ cutoffs ($K \in \{10, 20, 50, 100\}$) at which the difference
    was significant, from 1 to 4 (all $K$), so that brighter cells
    reflect increasingly robust spatial specificity. Empty grey cells
    indicate pairs that reached significance at no value of $K$, or for
    which either group was too small for testing.}
\vspace{15pt}
\label{fig:morphspec}
\end{figure}

Both groups contained sufficient samples for 94 tumor type–signature pairs, yielding 752 comparisons across two spatial-specificity metrics and four cutoff values (\(K \in \{10,20,50,100\}\)). At nominal significance (\(p<0.05\)), 41 pairs showed greater spatial specificity in the positive group in at least one analysis setting. These pairs spanned 22 tumor types and 15 signatures (Figure~\ref{fig:morphspec}). All nominally significant effects followed the prespecified direction: when a mutational process was genomically active, its attribution map showed lower overlap with the maps of other signatures or higher uniqueness. The Jaccard overlap identified 33 nominally significant pairs, whereas the uniqueness metric identified 26, with 18 pairs identified by both metrics. The patterns were also consistent across cutoff values. Of the 59 nominally significant pair–metric combinations, 34 were significant at two or more values of \(K\), and 14 were significant at all four.

To account for multiple comparisons, Benjamini--Hochberg correction
was subsequently applied separately within each metric and cutoff combination. Sixteen of the
41 tumor type--signature pairs retained an adjusted $q<0.05$ in at least one analysis setting.
These define a core subset of associations that remained supported after correction, while the
broader set of nominal associations was retained to assess recurrence and biological coherence
across tumor types and analysis settings.

Several of the observed tumor type--signature associations were consistent with established
biological relationships or previously reported tissue distributions. This concordance is
notable given the nature of the supervision. Hist2Sig receives no information about signature
identity or etiology during training. Its targets are simply per-sample exposure counts across
the 30 signatures, with no indication of which mutational processes these counts represent.
Among the associations supported after false discovery rate control, SBS4 in head and neck
squamous cell carcinoma (HNSC) was consistent with its tobacco-associated etiology; SBS10a in
uterine corpus endometrial carcinoma (UCEC) with its association with POLE exonuclease-domain
mutations; SBS44 in stomach adenocarcinoma (STAD) with defective DNA mismatch repair; and the
APOBEC-associated SBS2 in CESC and ESCA. SBS12 was supported in liver hepatocellular carcinoma
(LIHC), a tumor type in which this signature has previously been reported despite its uncertain
etiology. Additional nominal associations were similarly coherent. SBS44 extended to colon
adenocarcinoma (COAD) and UCEC, where mismatch repair--deficient tumors can exhibit prominent
lymphocytic infiltration, particularly in colorectal cancer~\cite{greenson2009pathologic};
SBS2 and SBS13 were observed in HNSC and BRCA, and SBS13 in ESCA; SBS7b in skin cutaneous
melanoma (SKCM), consistent with UV exposure; and SBS17b in STAD, consistent with its known
tissue distribution.

Signatures with incompletely characterized etiologies, showed a similar hierarchy of evidence. SBS40a remained significant
after correction in BRCA, KIRC, and LUSC, SBS40b in KIRC and KIRP, and SBS40c in KIRP;
together with the FDR-supported SBS44 association in STAD, the SBS40 family produced some of
the strongest and most recurrent signals in the analysis. Additional associations that did not
meet the adjusted threshold nevertheless recurred across tumor types or analysis settings.
SBS8 showed nominal spatial-specificity signals in seven tumor types, SBS18 in READ, SBS39 in
TGCT, THCA, and THYM, and the clock-like SBS1 in six tumor types. Although these recurrent
nominal patterns cannot be considered confirmatory, their consistency across tumor contexts
and analysis settings supports their prioritization as candidate morphology-associated signals
for validation in independent cohorts.

\begin{figure}[h!]
\centering

\begin{minipage}{0.83\linewidth}
\centering

\panel{a}{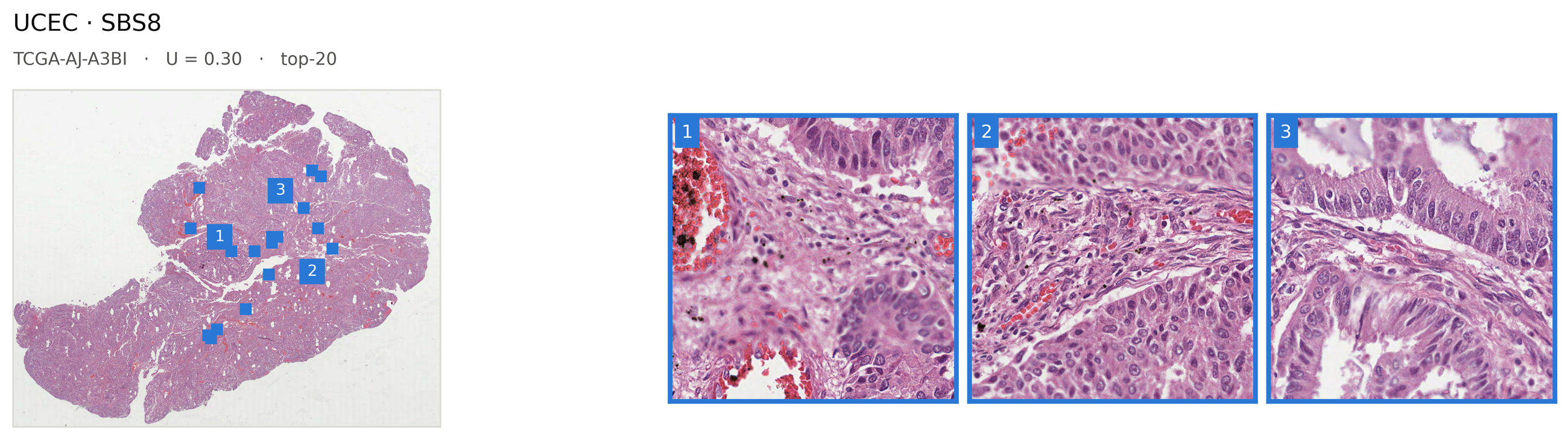}

\vspace{-0.9em}

\panel{b}{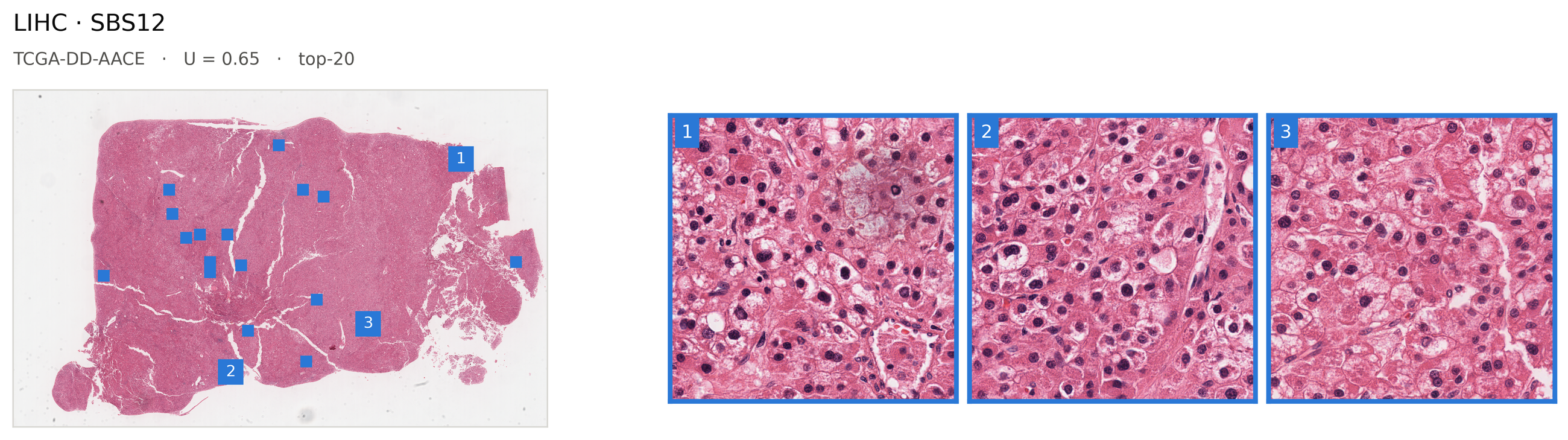}

\vspace{-0.9em}

\panel{c}{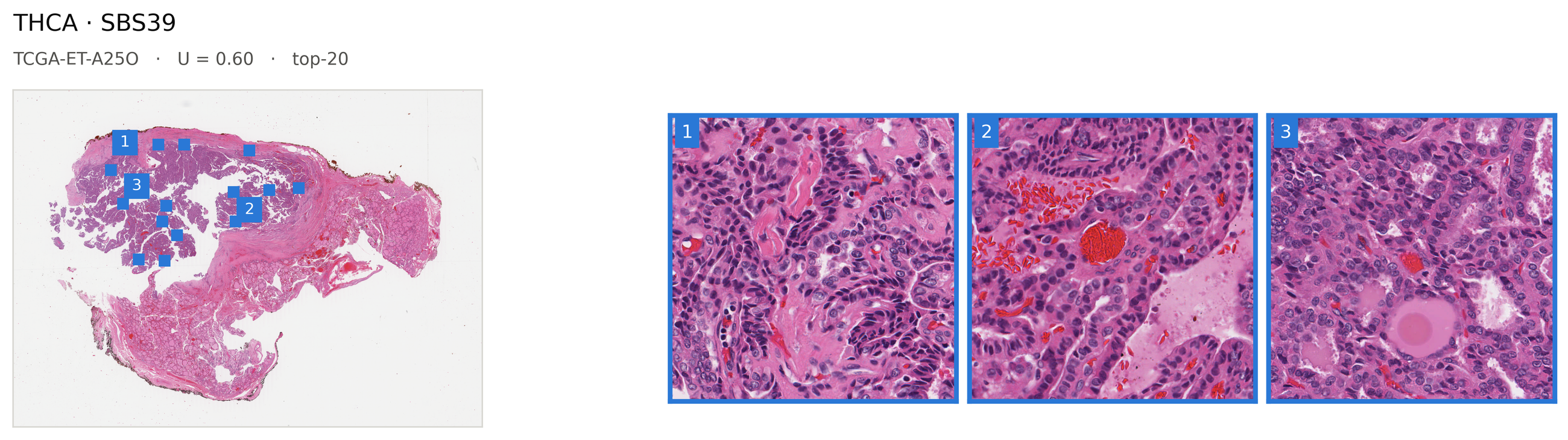}

\vspace{-0.9em}

\panel{d}{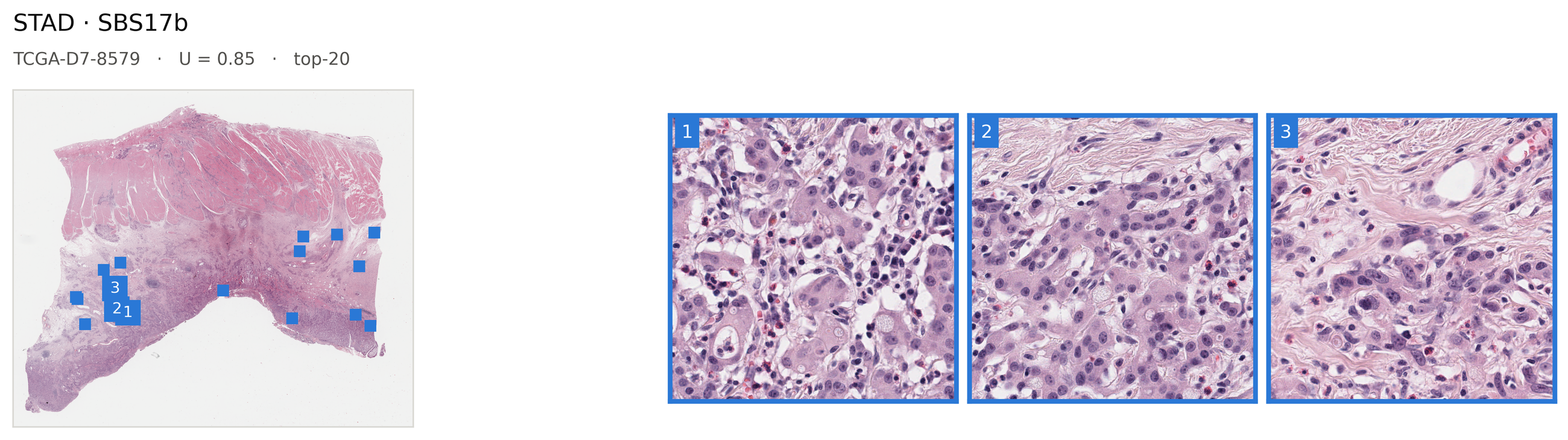}

\end{minipage}

\vspace{0.1em}

\caption{\textbf{Most strongly attributed patches for four signatures of
uncertain etiology.} For each panel, one true-positive slide is shown: the
signature is among the three largest exposures in the genomic profile and is
also among the three highest activities predicted by the model. Left,
whole-slide overview with the $K=20$ patches having the highest combined
attention--gradient score $c_i^{(k)}$; numbered markers correspond to the
enlarged patches on the right, ordered by decreasing score. $U$ is the
uniqueness $U_{20}$ defined in Section~\ref{subsec:morph_methods}, i.e. the
fraction of those 20 patches that do not appear in the top-20 of any other
signature co-dominant in the same slide.
\textbf{(a)} UCEC, SBS8;
\textbf{(b)} LIHC, SBS12;
\textbf{(c)} THCA, SBS39;
\textbf{(d)} STAD, SBS17b.}
\label{fig:patch_examples_unknown}
\end{figure}

Beyond aggregate predictive performance, this analysis provides a framework for generating and
prioritizing hypotheses about the morphological specificity of mutational processes. In
particular, it may reveal previously unrecognized histological features associated with
individual signatures and direct pathologists to the image regions in which these candidate
features are most visible. To facilitate this interpretation, we identify and visualize, for
each signature, the patches with the highest attribution scores.
Given the large number of tumor types and mutational signatures analyzed, the main text
presents selected examples focusing on signatures with unknown or uncertain etiologies, namely
SBS8, SBS12, SBS17b, and SBS39 (Figure~\ref{fig:patch_examples_unknown}). As a morphological
reference, we show the same visualization for four signatures whose mutational process is
established, namely SBS18, SBS44, SBS2, and SBS13
(Figure~\ref{fig:patch_examples_known}); additional examples are provided in the Supplementary
Materials. These visualizations should be interpreted as hypothesis-generating and will
require confirmation through expert pathological assessment and independent spatial or
molecular validation.

\begin{figure}[h!]
\centering

\begin{minipage}{0.83\linewidth}
\centering

\panel{a}{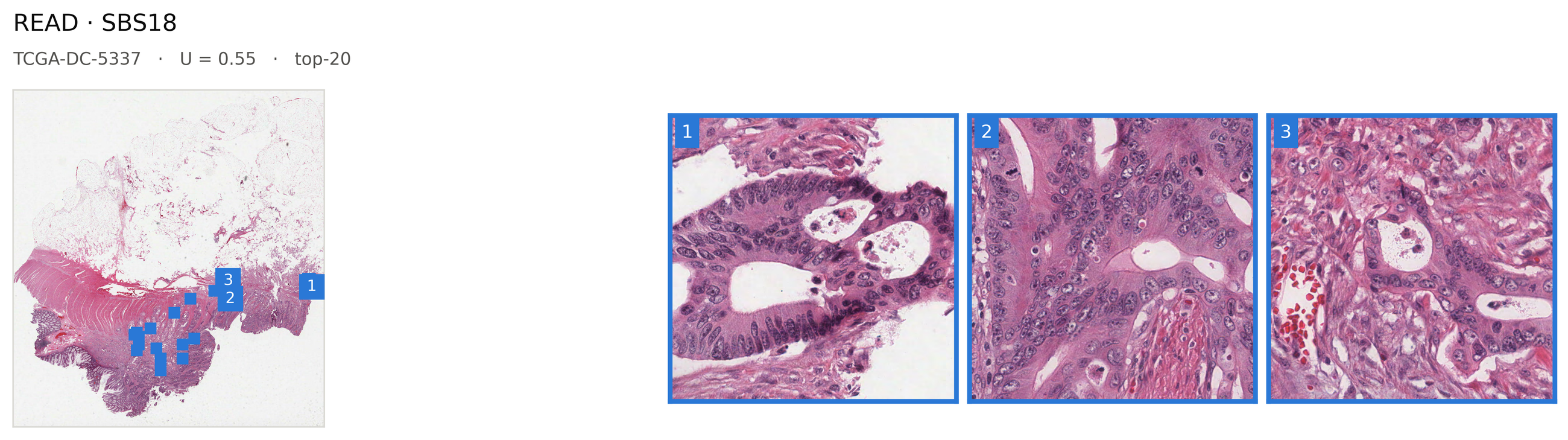}

\vspace{-0.9em}

\panel{b}{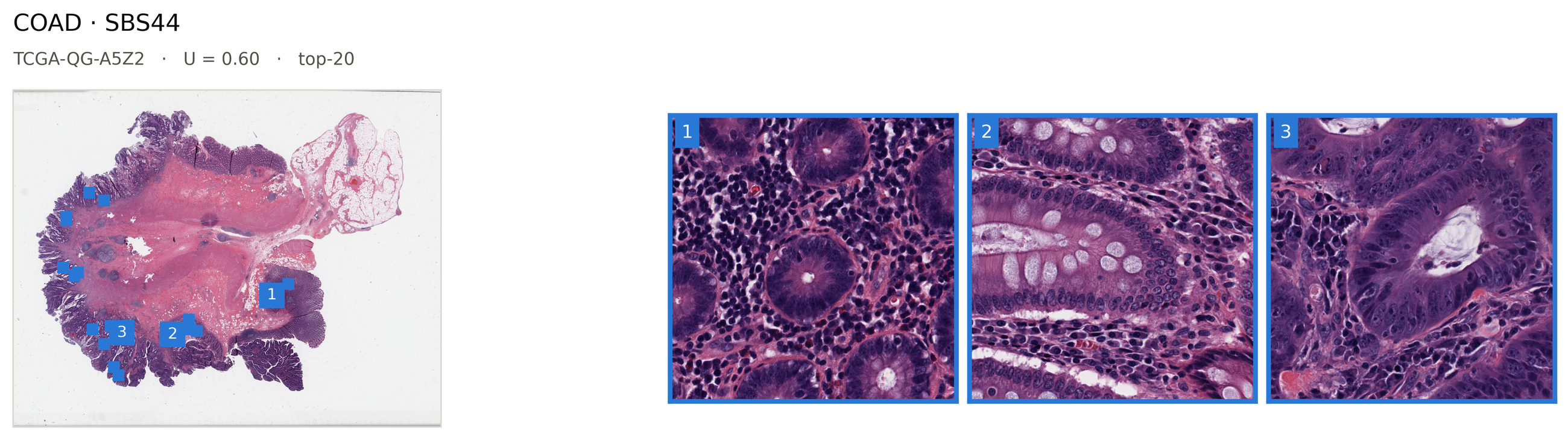}

\vspace{-0.9em}

\panel{c}{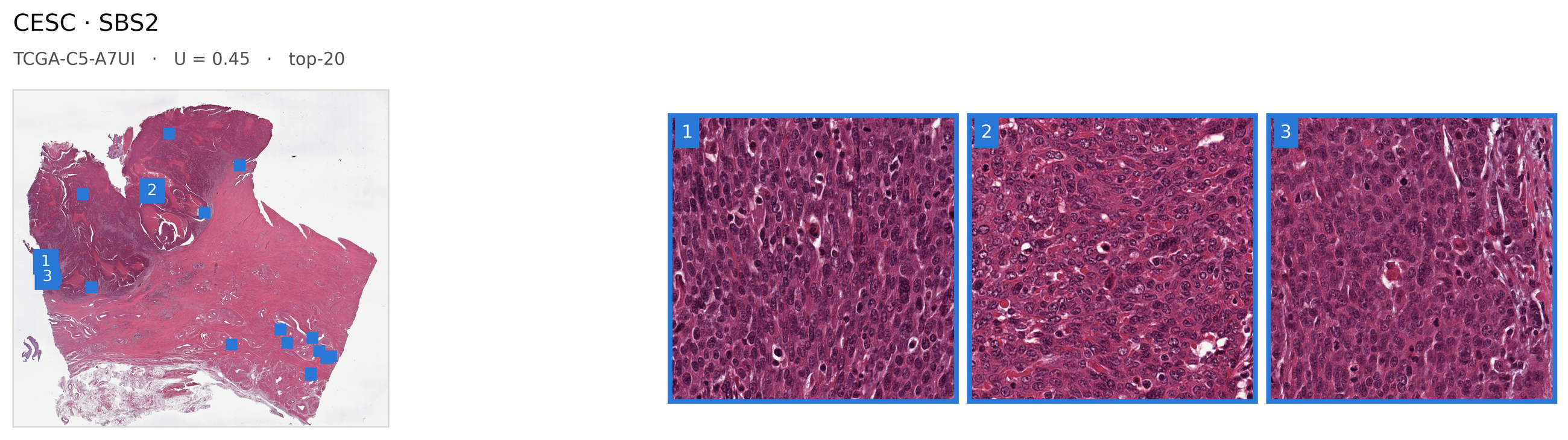}

\vspace{-0.9em}

\panel{d}{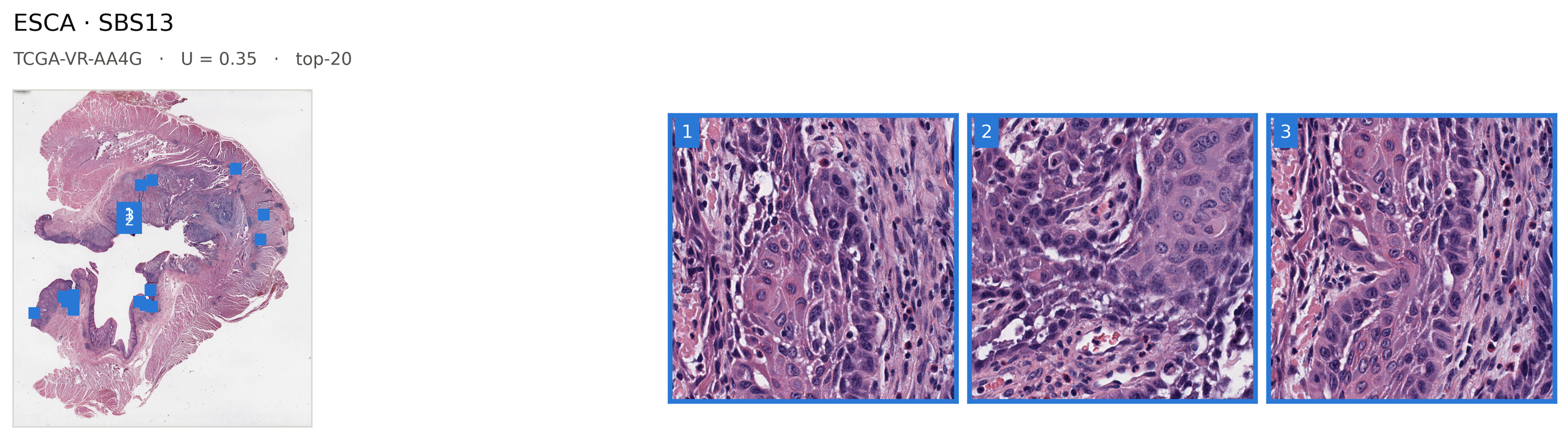}

\end{minipage}

\vspace{0.1em}

\caption{\textbf{Most strongly attributed patches for four signatures of
established etiology.} Same layout and selection criterion as
Figure~\ref{fig:patch_examples_unknown}: one true-positive slide per panel,
whole-slide overview with the $K=20$ highest-scoring patches, and the three
top-scoring patches enlarged on the right. These signatures act as a
morphological reference, since the underlying mutational process is known:
SBS18 reflects damage by reactive oxygen species, SBS44 defective mismatch
repair, and SBS2 and SBS13 APOBEC cytidine deaminase activity.
\textbf{(a)} READ, SBS18;
\textbf{(b)} COAD, SBS44;
\textbf{(c)} CESC, SBS2;
\textbf{(d)} ESCA, SBS13.}
\label{fig:patch_examples_known}
\end{figure}

\subsection{Hist2Sig evaluation on CPTAC cohort}\label{subsec:cptac}
Hist2Sig results on the TCGA cohort show that inferring mutational signature exposures from WSI images is feasible, although the morphological signal of signatures varies across tumor types. To assess the generalizability of these findings, we evaluated the model on the CPTAC cohort. After genomic filtering, Mutect2-based sample selection, and WSI matching, the external evaluation cohort comprised 193 samples across five tumor types (see Section \ref{mutsig}). All five tumor types have a direct counterpart in the TCGA training cohort (KIRC, PAAD, LUAD, GBM, and LUSC), ensuring that external evaluation is performed on tissues of origin already represented during training. Nevertheless, TCGA and CPTAC represent distinct cohorts and differ in several aspects, including tissue processing protocols, staining procedures, and scanner hardware.

Although the relative signature compositions extracted by SigProfilerAssignment are highly similar between the two cohorts (Supplementary Figure S5), substantial differences emerge in the absolute mutation load, defined as the sum of signature exposure counts. Supplementary Figure S6 compares the observed and predicted mutation loads across the five matched tumor types in TCGA and CPTAC. In the observed genomic data, TCGA samples consistently show a higher mutation load than their CPTAC counterparts across all five tumor types, with the largest differences observed in lung adenocarcinoma and lung squamous cell carcinoma (LUAD and LUSC). Thus, despite the broadly conserved relative signature composition, the two cohorts exhibit a systematic shift in the absolute scale of the genomic target.

This cohort-level difference is also reflected in the predictions produced by Hist2Sig. The relative ranking of tumor types by mutation load is largely preserved, with LUSC and LUAD showing the highest loads and KIRC and PAAD the lowest, and the direction of the TCGA--CPTAC difference is reproduced across tumor types. However, the model amplifies the difference in absolute scale. Predicted mutation loads in CPTAC tend to be compressed toward lower values, whereas in TCGA the model tends to overestimate the observed loads. This effect is particularly evident for GBM, where predicted counts in TCGA are markedly higher than the corresponding observed values. Hist2Sig therefore captures the overall structure of the differences between cohorts and tumor types, but shows imperfect calibration of absolute signature exposures when transferred to the external cohort.

Importantly, the difference between TCGA and CPTAC is not restricted to their genomic profiles but is also evident in the histological data that constitute the actual input to Hist2Sig. To characterize this input-domain shift, we examined the slide-level embeddings generated by the H-optimus-v1 foundation model and used as input to Hist2Sig. The embeddings show a clear separation between TCGA and CPTAC, both in the joint representation space and within individual tumor types (Supplementary Figure S7). The persistence of cohort separation within matched tumor types indicates that this shift cannot be explained solely by differences in tissue-of-origin composition and instead reflects systematic differences between the histological representations of the two cohorts.

This observation is particularly relevant to the external evaluation setting. Hist2Sig is trained exclusively on TCGA, where it learns a mapping from histological features to mutational signature exposures. At inference time, this trained mapping is applied to CPTAC using histological images alone; CPTAC genomic data are not provided to the model and are used only as ground truth to evaluate its predictions. The model is therefore required to transfer an image-to-genomics relationship learned entirely from TCGA to a cohort whose histological feature distribution differs systematically from that encountered during training. At the same time, the ground-truth genomic distribution in CPTAC also differs from TCGA in its absolute mutation load, despite the relative signature compositions remaining broadly similar. The external evaluation therefore combines an input-domain shift in histology with a shift in the distribution of the genomic quantities that the model is expected to predict.

These two sources of cohort heterogeneity provide important context for the observed calibration differences. Technical factors such as tissue preparation, staining, and scanner hardware may contribute to the shift in histological representations, while biological, demographic, and other cohort-specific differences may contribute to variation in both histological and genomic profiles. Our analyses do not establish which of these factors is responsible for the observed prediction shift, nor do they demonstrate a causal relationship between the histological domain shift and the differences in predicted mutation load. Nevertheless, the simultaneous presence of a substantial input-domain shift and a difference in the absolute genomic target distribution highlights the difficulty of transferring a mapping learned exclusively on TCGA to CPTAC and provides a plausible explanation for the reduced calibration of Hist2Sig predictions on the absolute exposure scale.

Despite these cohort-level differences, the presence of domain shift does not by itself determine whether the relationship between histological features and mutational signature profiles learned on TCGA transfers to CPTAC. We therefore next assessed the predictive performance of Hist2Sig on the external cohort. Given the systematic differences in absolute mutation load described above, we first focused on cosine similarity, which is insensitive to the overall magnitude of the exposure vectors and instead measures how well the model recovers their relative signature composition.

\begin{figure}[h!]
\centering
\includegraphics[width=.75\linewidth]{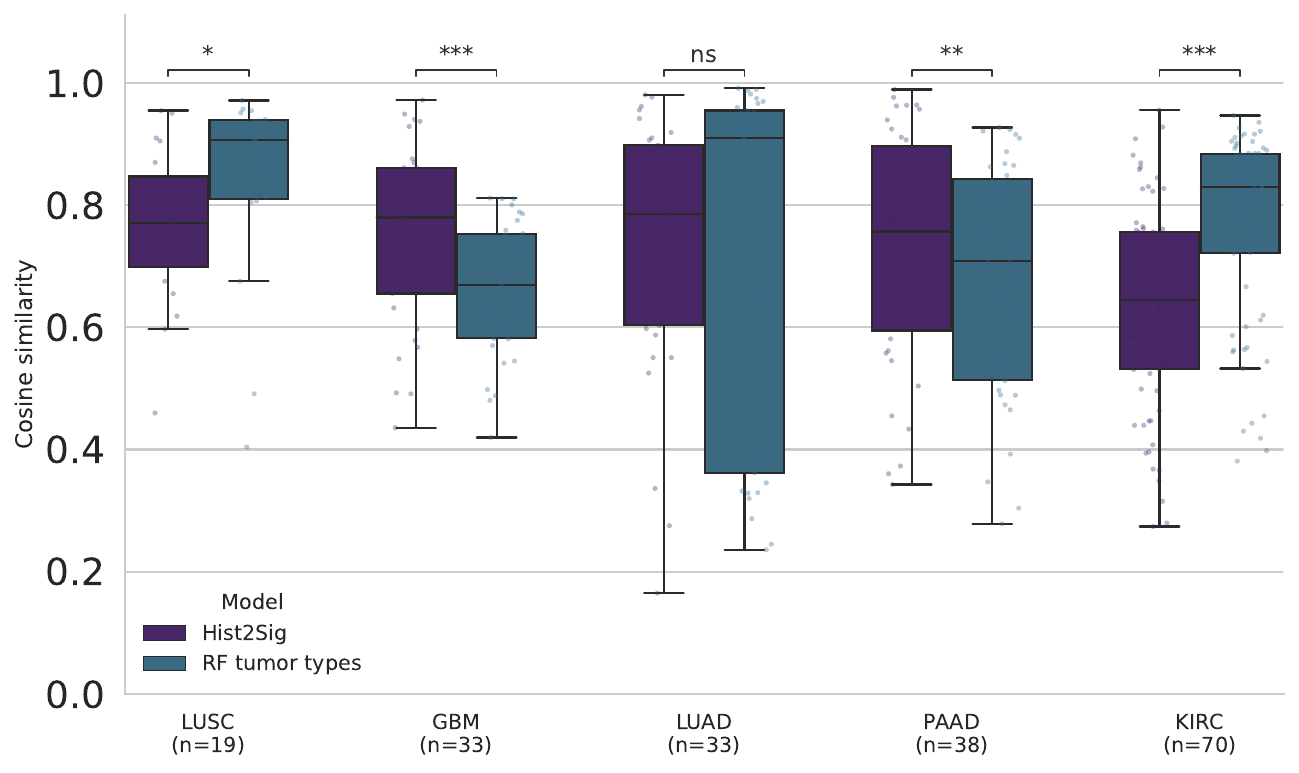}
\caption{\textbf{Comparison of cosine similarity distribution in the CPTAC cohort}. The cosine similarity is computed between predicted and ground-truth mutational signature exposure profiles across tumor types. Box plots show the median and interquartile range. Hist2Sig is represented in violet while the Random Forest in blue. Asterisks denote statistically significant differences between the two models (Wilcoxon signed-rank test; $* p < 0.05$, $** p < 0.01$, $*** p < 0.001$)}
    \label{fig:cosine_cptac}
\end{figure}

Figure~\ref{fig:cosine_cptac} reports the cosine similarity distributions on the CPTAC cohort for Hist2Sig and the Random Forest (RF) baseline. Hist2Sig achieves median cosine similarities between 0.65 and 0.78, indicating that, despite the shift in both histological representations and absolute mutation load, the model retains some ability to recover the relative mutational signature composition in an external cohort not used during training. LUSC, GBM, LUAD, and PAAD reach median values of approximately 0.77--0.78, whereas KIRC shows lower concordance. Compared with the tumor type--only baseline, Hist2Sig performs significantly better in GBM and PAAD, whereas the RF baseline performs significantly better in LUSC and KIRC; no significant difference is observed in LUAD.

\begin{figure}[h!] 
\centering \includegraphics[width=1\linewidth]{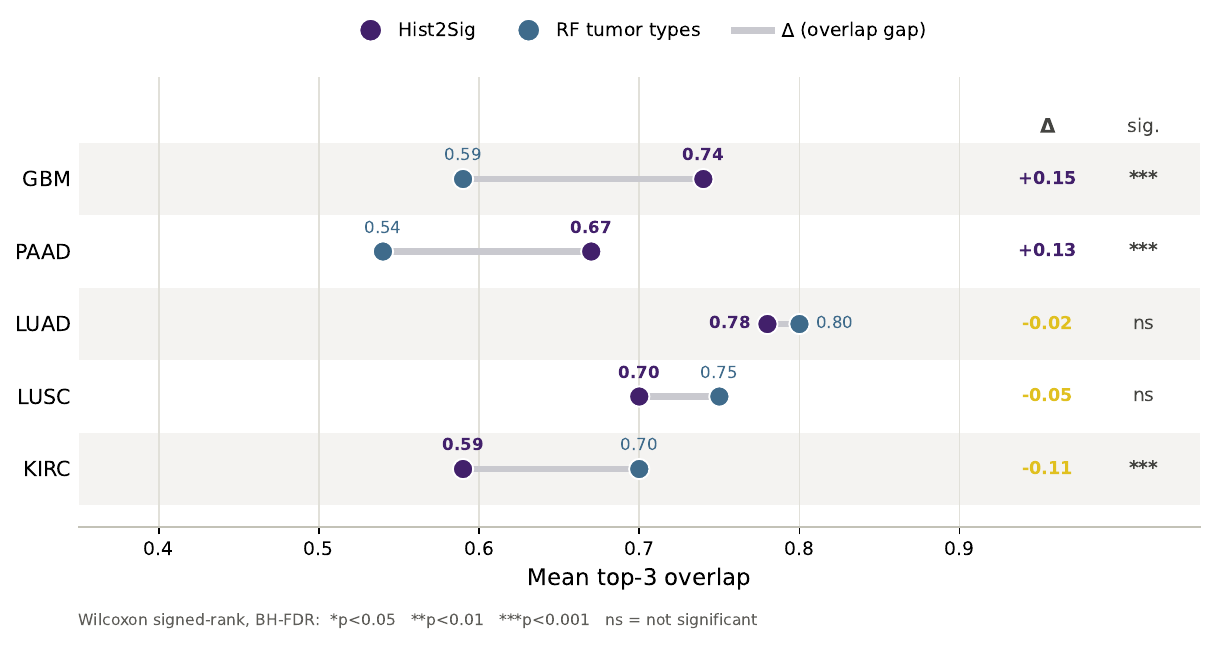}
\caption{\textbf{Mean top-3 signature overlap on the CPTAC cohort.} For
    each tumor type, the two dots show the mean fraction of correctly
    identified dominant signatures for Hist2Sig (violet) and a Random
    Forest baseline trained on tumor type labels (blue); the grey
    connector spans the gap between them. The $\Delta$ column reports the
    pairwise difference in mean overlap (Hist2Sig $-$ RF), coloured
    purple where Hist2Sig outperforms the baseline and red where the
    baseline prevails. Statistical significance of the difference between
    the two models is indicated in the adjacent column (Wilcoxon
    signed-rank test, Benjamini--Hochberg FDR correction;
    $\ast p < 0.05$, $\ast\ast p < 0.01$, $\ast\ast\ast p < 0.001$; ns,
    not significant). \label{fig:overlap_cptac}}
\end{figure}

To better evaluate whether Hist2Sig captures morphological information beyond
what can be inferred from the tumor type label alone, we again compared its
mean top 3 overlap score with the Random Forest baseline (Figure~\ref{fig:overlap_cptac}). In agreement with Figure \ref{fig:cosine_cptac} in GBM and PAAD,
Hist2Sig substantially outperforms the baseline ($+0.15$ and $+0.13$,
respectively), indicating that the model extracts morphological features
that carry signature level information not reducible to tissue of origin.
In LUAD, LUSC and KIRC the baseline performs on par or slightly better
than Hist2Sig ($-0.02$, $-0.05$, and $-0.11$). For the latter tumor types,
the dominant mutational processes are strongly tissue-associated (e.g.\
SBS4 in lung cancers, linked to tobacco exposure), meaning that knowing
the tumor type already provides a strong prior on the signature profile.
In this regime, the baseline benefits from a deterministic mapping between tissue label and dominant signatures, whereas Hist2Sig must
recover this information from morphology alone, a harder task that nonetheless yields competitive overlap scores.

\begin{figure}[h!]
  \centering
\includegraphics[width=0.8\linewidth]{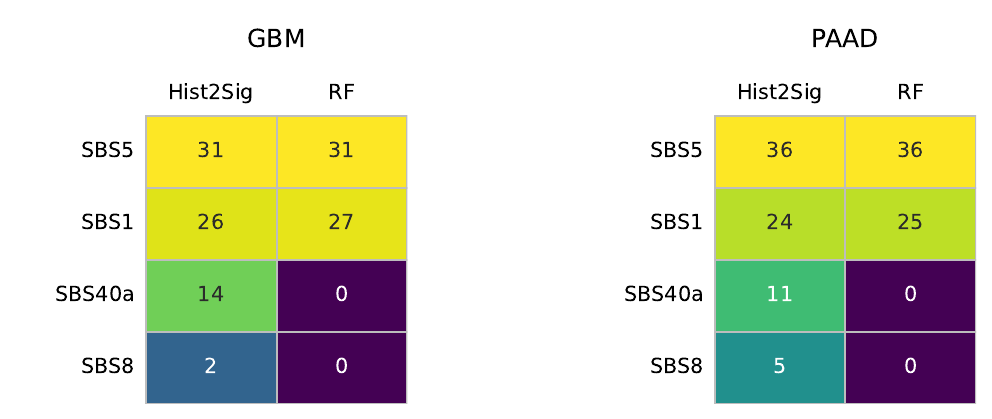}
  \caption{\textbf{Signature diversity in Top 3 predictions for GBM and PAAD on the CPTAC cohort.} Cell values indicate the number of samples in which each signature was correctly identified among the top three by Hist2Sig and by the Random Forest baseline (RF); color intensity (log scale) reflects the magnitude of the count.}
  \label{fig:diversity_cptac}
\end{figure}

Interestingly, in GBM and PAAD, the two tumor types where Hist2Sig most
clearly outperforms the tumor type-only baseline, signatures SBS8 and
SBS40a are frequently predicted as dominant, whereas the Random Forest fails to identify them (Figure \ref{fig:diversity_cptac}).
Both signatures were also correctly recovered
in the TCGA cohort. The fact that Hist2Sig recovers these signatures as dominant from histology on a very different external cohort suggests that the network may be leveraging morphological correlates of
these mutagenic processes.
Taken together, these results indicate that the relationship learned by Hist2Sig on TCGA transfers, at least in part, to an independent cohort acquired under different staining, scanning and variant calling protocols. The transfer is most informative precisely where it matters for the central question of this study. In GBM and PAAD, the tumor types where the TCGA analysis had already shown that morphology carries information beyond tissue of origin, Hist2Sig reproduces the same advantage over the tumor type baseline and recovers the same signatures that are not simply dictated by tissue identity, namely SBS8 and SBS40a, despite the substantial shift in the underlying histological representations documented in Supplementary Figure S7. Conversely, in LUAD, LUSC and KIRC, where the TCGA results indicated that tumor type alone is close to sufficient, the external comparison reproduces the same pattern, with the baseline matching or slightly exceeding Hist2Sig. This agreement between the internal and external cohorts, obtained despite a documented shift in the histological input and in the absolute genomic target distribution, supports the interpretation that the signal exploited by Hist2Sig reflects genuine associations between histology and genotype rather than artifacts specific to one cohort. This holds even though, as shown above, the absolute calibration of predicted exposures does not transfer equally well.

\newpage
\section{Discussion}

We asked whether mutational signature exposures can be inferred from routine H\&E slides at pan-cancer scale and, more specifically, whether any predictive signal reflects morphology beyond tissue of origin. Hist2Sig, trained on 7,063 whole-genome-sequenced tumors spanning 29 tumor types, recovered relative signature compositions with a median cosine similarity above 0.8 in most tumor types and identified the dominant mutational processes more accurately than a tumor-type-only baseline in 19 of the 29. The task is therefore feasible, although performance varies substantially across tumor types and signatures. This variation is itself informative because it indicates where morphology contributes information and where it does not.

When recovering the three dominant mutational signatures, Hist2Sig exceeded the tumor-type-only baseline by up to 0.30 in top-three overlap for COAD, 0.20 for PRAD, 0.18 for GBM, and 0.17 for UCEC. Its largest deficit in any tumor type was only 0.06. Where the baseline performed better, it did so by small margins, even when the differences were statistically significant. For signatures strongly associated with tissue of origin, such as SBS4 in lung cancer and SBS7a/b in melanoma, the tumor-type label already captures much of the available predictive information, leaving limited scope for morphology to improve performance. In tumors with more heterogeneous mutational landscapes, however, histology appears to resolve variation between patients that cannot be captured by tissue type alone.

The signatures recovered by Hist2Sig further support this interpretation. SBS1, SBS8, SBS18, SBS39, and SBS40a were repeatedly identified among the dominant processes in cases where the tumor-type-only baseline failed. At the individual-sample level, some of the strongest correlations involved signatures associated with biological processes that could plausibly produce recognizable morphological phenotypes, including the APOBEC-associated SBS2 and SBS13, the clock-like SBS1, and the mismatch-repair-associated SBS44. The model was not provided with explicit information about the etiologies of these signatures or their possible histological manifestations; it received only per-sample exposure counts for the 30 modeled signatures. Nevertheless, signature identity and tissue distribution remain implicit in the training targets, and these results alone cannot establish that the model recognizes the underlying mutational processes. Rather, they are consistent with Hist2Sig detecting morphological phenotypes associated with those processes.

The spatial attribution analysis provides complementary evidence. Across 94 tumor type--signature pairs, patches attributed to a given signature became more spatially distinct from patches attributed to co-occurring signatures when that signature was genomically dominant. This association reached nominal significance in 41 pairs, 16 of which retained $q < 0.05$ after Benjamini--Hochberg correction within each metric and cutoff. Several of these pairs correspond to established biological associations that were not supplied through spatial supervision, including SBS4 in HNSC, SBS10a in UCEC, and SBS44 in STAD. These findings suggest that the model does not rely exclusively on a single, generically informative tissue compartment but instead associates different signatures with distinct morphological regions. The remaining nominal associations recurred across tumor types and cutoffs but should be regarded as candidates for independent validation rather than as definitive findings.

This spatial structure also suggests a role for Hist2Sig beyond prediction. By ranking the patches that contribute most strongly to each signature, the model highlights regions in which potential morphological correlates may be most apparent. This is particularly relevant for SBS8, SBS12, and SBS39, whose etiologies remain uncertain, and for SBS18, which has been linked to damage by reactive oxygen species but lacks a well-established histological correlate. The resulting maps are hypothesis-generating rather than confirmatory, but they provide a tractable starting point for expert pathological review and targeted molecular or spatial assays.

External evaluation in CPTAC showed that the predictive signal generalizes, at least in part, to a cohort generated using different tissue-processing, staining, scanning, and variant-calling protocols. Hist2Sig achieved median cosine similarities ranging from 0.65 to 0.78 and retained a clear advantage over the baseline in GBM ($+0.15$) and PAAD ($+0.13$), the same tumor types in which the TCGA analysis indicated that morphology provides information beyond tissue of origin. It also recovered SBS8 and SBS40a as dominant signatures in cases where the baseline did not. Conversely, in LUAD, LUSC, and KIRC, where the TCGA results suggested that tumor type alone was nearly sufficient, the baseline again matched or slightly outperformed the model. The agreement between the internal and external cohorts regarding where morphology adds information, despite a documented shift in histological representation space, makes it less likely that the observed signal is specific to TCGA.

Generalization was nevertheless incomplete in terms of absolute exposure values. Relative to the observed exposures, predictions were shifted toward lower values in CPTAC and were overestimated in TCGA, most visibly in GBM. Differences in mutation burden, cohort composition, and variant-calling procedures may all contribute to this discrepancy, as may the use of a zero-inflated negative binomial objective for highly heterogeneous count data. Their relative contributions remain to be determined. We deliberately modeled absolute exposures rather than compositional proportions because normalization removes information about overall mutation burden: the same relative contribution may represent thousands of mutations in a hypermutated tumor but only a handful in a low-burden sample, where the estimate may be close to the noise floor. Converting exposures to proportions would therefore conceal the calibration problem rather than resolve it. At present, Hist2Sig predictions are better suited to ranking samples and comparing relative attribution patterns than to estimating absolute exposure counts. Developing calibration methods that remain valid under cohort shift is therefore an important priority.

Several limitations constrain these conclusions. First, a representation shared across 29 tumor types may dilute disease-specific patterns and underrepresent tumor types with small sample sizes or unusual mutational landscapes. Tumor-specific models trained on larger cohorts may therefore capture associations missed by the pan-cancer model. Second, the supervision is spatially coarse: a single exposure vector derived from bulk sequencing is assigned to an entire slide, even though both morphology and mutational activity may vary across regions. Third, the external cohort includes only five tumor types with modest sample sizes, limiting the precision with which calibration and signature-specific performance can be assessed. Fourth, the targets are exposure estimates generated by SigProfilerAssignment rather than direct ground truth. They depend on the reference signature catalog, upstream variant-calling procedures, and the reconstruction-similarity threshold; any systematic errors introduced during this process may therefore be inherited by the model. Finally, the analysis was restricted to SNV-derived SBS signatures and excluded indel and structural-variant signatures, which capture complementary aspects of processes such as homologous recombination deficiency.

Although its immediate clinical applicability remains limited, this feasibility study demonstrates that mutational signature exposures can be inferred from routine histology and that, for several tumor types, morphology provides information beyond tissue of origin. These findings establish a foundation for developing larger, tumor-type-specific models that may capture disease-specific associations more effectively. With further validation and appropriate calibration, such models could support clinically relevant patient stratification and help distinguish among co-occurring mutational processes, including identifying which process predominates in an individual tumor. The attribution maps may also guide the discovery and pathological characterization of morphological phenotypes associated with these processes. Prospective evaluation in independent cohorts, together with expert review of the highlighted regions and comparison with molecular measurements, will be required to determine whether these potential applications translate into clinical value.

\section{Methods}\label{sec2}

\subsection{Data acquisition and feature extraction}
\label{Data preprocessing}

We obtained Histopathological whole-slide images (WSIs) and matched whole-genome sequencing (WGS) data from The Cancer Genome Atlas (TCGA) via the Genomic Data Commons (GDC) portal (accessed May 15, 2025). Access to controlled WGS data was granted upon request and approved by the GDC Data Access Committee.

The initial TCGA dataset included $10{,}330$ patients across 32 primary tumor types with available diagnostic WSIs and corresponding somatic variant call format (VCF) files; after signature decomposition, filtering, and matching with histopathology data, the final modeling cohort comprised 7,063 patients across 29 tumor types. For this study, we focused exclusively on somatic single-nucleotide variants (SNVs), which are derived from the GATK \texttt{Mutect2} pipeline~\cite{van2020genomics}.

An independent external cohort was obtained from the Clinical Proteomic Tumor Analysis Consortium (CPTAC) through the GDC portal upon request. This dataset includes 1,340 samples across 11 tumor types with available whole-genome sequencing (WGS) data. The same preprocessing and feature extraction pipeline used for TCGA was applied to CPTAC to ensure methodological consistency and enable external validation.

One important difference lies in the somatic mutation calling pipelines used for the two datasets from the GDC portal. In TCGA, annotated somatic mutations in VCF files are called using GATK \texttt{Mutect2}, whereas CPTAC samples are mainly processed with VarScan2 \cite{koboldt2012varscan}. For consistency, we retained only CPTAC samples whose somatic mutations were annotated using GATK \texttt{Mutect2}, resulting in 193 samples with both H\&E images and matched somatic variant data.

The majority of WSIs were originally scanned at 40$\times$ magnification. To balance computational efficiency with tissue detail, we downsampled images to an effective 20$\times$ resolution using the \texttt{pyvips} Python library. Patches were extracted from each slide using an in-house pipeline adapted from CLAM~\cite{lu2021data}, and we performed stain normalization  using the Vahadane method~\cite{vahadane2016structure} to minimize variability.

We then extracted high-dimensional visual features  using \texttt{H-optimus-v1}~\cite{hoptimus1}, a 1.1B-parameter Vision Transformer trained in a self-supervised manner on a proprietary dataset of billions of histology tiles from over one million slides and more than 800,000 patients. \texttt{H-optimus-v1} has shown state-of-the-art performance on several computational pathology benchmarks~\cite{campanella2025clinical,breen2025comprehensive}. For each WSI, the model produced a feature matrix of size $N \times 1536$, where $N$ corresponds to the number of extracted patches.
\subsubsection{Mutational signature analysis}\label{subsec21}

We analyzed somatic variants in VCF format using \texttt{SigProfilerAssignment}~\cite{diaz2023assigning} to decompose each sample's mutational profile into exposures to known single base substitution (SBS) signatures. We then excluded sequencing artifacts and signatures linked to contamination according to COSMIC guidelines~\cite{sondka2024cosmic}.

This analysis was first applied to the TCGA cohort and then replicated in the CPTAC cohort. Both datasets underwent identical procedures for signature decomposition, filtering, and post-processing. To ensure robust analysis while preserving tumor type diversity, two filtering steps were applied: 

\begin{enumerate}
    \item Only samples with a reconstruction similarity of at least $0.9$ (between original and reconstructed mutational profiles) were retained, as reported by \texttt{SigProfilerAssignment}.
    \item Tumor types represented by fewer than 50 samples after
    signature decomposition were excluded. Following this filtering step,
    samples were matched with their corresponding histopathological WSIs
    available in the GDC portal. As a result of this matching, some tumor
    types may contain fewer than 50 samples in the final dataset, as
    not all genomically filtered samples have a paired WSI available.
    
\end{enumerate}

For each patient, the five SBS signatures with the highest exposure were selected. Then, the union of these top signatures across all patients was computed, and any signature present in fewer than 50 samples was removed to ensure sufficient representation for downstream analyses. 

\subsubsection{Hist2Sig architecture}

We obtained the final dataset combining the high-dimensional visual features extracted with H-optimus-v1 and the mutational signature exposures information. For each patient, we associated the available histopathological WSI features with a single vector of exposure counts corresponding to the 30 retained mutational signatures; when multiple WSIs were available for the same patient, all slides were kept within the same cross-validation fold.
Hist2Sig is an attention-based Multiple Instance Learning (MIL) framework designed to predict mutational signature exposures from whole-slide image features.
To ensure efficient and robust training, we adopt a composite loss function composed of multiple complementary terms. Since mutational signature exposures are sparse and overdispersed count data, we employ a Zero-Inflated Negative Binomial (ZINB) loss as the main optimization objective:
\begin{equation}
\mathcal{L}_{\text{ZINB}} = \text{ZINB}(E, \hat{E})
\end{equation}
where $E \in \mathbb{R}^{30}$ denotes the ground-truth exposure vector and $\hat{E}$ the predicted exposures.
To enforce biological consistency between predicted exposures and the original mutational profiles, we introduce a reconstruction constraint. Let:
\begin{equation}
M \approx S E
\end{equation}
where $M \in \mathbb{R}^{d}$ is the observed mutational profile, $S \in \mathbb{R}^{d \times 30}$ is the mutational signature matrix, and $E \in \mathbb{R}^{30}$ is the exposure vector.
Given the predicted exposures $\hat{E}$, we reconstruct the mutational profile as:
\begin{equation}
\hat{M} \approx S \hat{E}
\end{equation}
Thus, we minimize the discrepancy between the original and reconstructed profiles using:
\begin{equation}
\mathcal{L}_{\text{rec}} = \left\| M - \hat{M} \right\|_2^2
\end{equation}
To help the model correctly identify the dominant mutational processes, we incorporate a SoftRank loss over the top-$k$ signatures (with $k=3$):
\begin{equation}
\mathcal{L}_{\text{rank}} = \text{SoftRank}_k(E, \hat{E})
\end{equation}
The overall objective function is:
\begin{equation}
\mathcal{L}_{\text{total}} =
\lambda_1 \mathcal{L}_{\text{ZINB}} +
\lambda_2 \mathcal{L}_{\text{rec}} +
\lambda_3 \mathcal{L}_{\text{rank}}
\end{equation}
where $\lambda_1$, $\lambda_2$, and $\lambda_3$ are weighting coefficients, whose values are chosen in the fine-tuning process.
The model was optimized on the TCGA cohort and internally evaluated using patient-level 5-fold cross-validation stratified by tumor type, ensuring that all slides from the same patient were assigned to the same fold; the final model was then evaluated externally on the CPTAC cohort.

\subsection{Evaluation of Mutational Signature Exposure Predictions}\label{sig_eval}

We evaluated Hist2Sig using multiple complementary strategies:

\begin{enumerate}

    \item \textbf{Exposure profile similarity.}
    For each tumor type, we computed the cosine similarity between predicted and
    ground-truth exposure profiles and report the median and interquartile range
    across samples. Because cosine similarity is invariant to vector magnitude,
    this metric evaluates whether the model correctly recovers the relative
    proportions of mutational signatures.

    \item \textbf{Recovery of dominant mutational signatures.}
    Since identifying dominant mutational processes is often clinically relevant,
    we assessed whether Hist2Sig correctly recovered the three signatures with highest exposure per sample. For each tumor type independently, we computed an
    unordered overlap score between the top three predicted and top three
    ground-truth signatures, without penalizing differences in ranking. The score
    corresponds to the fraction of correctly identified signatures among the top
    three and can take values of 0, 1/3, 2/3, or 1.

    \item \textbf{Comparison against a tumor-type baseline.}
    To contextualize model performance, we compared Hist2Sig against a Random
    Forest baseline trained using tumor type labels as input features. Because this
    baseline assigns identical exposure profiles to all samples within a tumor type,
    it captures how much of the mutational signature landscape can be explained by
    tumor type alone.

    \item \textbf{Assessment of intra-tumor-type heterogeneity.}
    We evaluated whether Hist2Sig captures sample-level heterogeneity
    within tumor types using two complementary analyses. First, we quantified the
    diversity of dominant predicted signatures across samples within each tumor type
    and compared it against the Random Forest baseline. Since the baseline assigns
    identical predictions within tumor types, any increase in diversity reflects
    sensitivity to intra-tumor variation.
    Second, for each tumor type and mutational signature, we computed the Pearson
    correlation coefficient between predicted and ground-truth exposures across
    samples, evaluating the ability of the model to resolve sample-level variation.
\end{enumerate}

\subsection {Assessing morphological specificity of signatures}\label{subsec:morph_methods}

To interpret the morphological evidence underlying signature predictions,
we derived per-patch attribution scores in a signature-specific manner,
without retraining. The procedure adapts a Grad-CAM-style formulation
\cite{selvaraju2017grad} to the MIL setting, combining the shared
attention weights with gradient information specific to each signature
output.

For a given slide, we performed a forward pass through Hist2Sig to obtain
the per-patch hidden embeddings $\mathbf{h}_i$, the attention weights
$a_i$, and the predicted exposure vector
$\hat{E} = (\hat{E}_1, \dots, \hat{E}_{30})$. For each signature $k$, we
computed the gradient of the corresponding logit with respect to the
hidden embeddings,
\begin{equation}
\mathbf{g}_i^{(k)} = \frac{\partial \hat{E}_k}{\partial \mathbf{h}_i},
\end{equation}
and defined the signed contribution of patch $i$ to signature $k$ as the
element-wise product summed over the embedding dimension,
\begin{equation}
s_i^{(k)} = \mathbf{h}_i \cdot \mathbf{g}_i^{(k)}.
\end{equation}
Following the Grad-CAM rationale, we retained the positive contributions
through a rectifier and modulated them by the MIL attention weight, which
encodes the relevance of each patch to the slide-level decision,
\begin{equation}
c_i^{(k)} = a_i \cdot \mathrm{ReLU}\!\left(s_i^{(k)}\right).
\end{equation}
The score $c_i^{(k)}$ therefore highlights patches that are both
attended-to by the model and whose embedding contributes positively to
the prediction of signature $k$. We computed $c_i^{(k)}$ for all 30 signatures. Beyond visualizing individual attribution maps, we sought to quantify whether
the morphological regions highlighted for a given signature are specific to that
signature, or whether they are largely shared with other co-occurring signatures.
To this end, we used two complementary statistics based on the patch-level attribution scores $c_i^{(k)}$: the \emph{Jaccard overlap}, which quantifies the pairwise overlap between the patches attributed to two signatures, and the \emph{uniqueness}, which quantifies the extent to which the patches attributed to a given signature are not shared with any of the other signatures. Let $\mathcal{T}_K(k)$ denote the set of the top-$K$ patches ranked by $c_i^{(k)}$. Specifically, we considered:

\begin{itemize}
\item \textbf{Jaccard overlap.} For a signature of interest $k$ and a co-occurring signature $l$, we define
\begin{equation}
J_K(k,l) =
\frac{\left| \mathcal{T}_K(k) \cap \mathcal{T}_K(l) \right|}
{\left| \mathcal{T}_K(k) \cup \mathcal{T}_K(l) \right|}.
\end{equation}
A low $J_K(k,l)$ indicates that the most strongly attributed patches of the two signatures are largely disjoint, whereas a high value indicates that the two signatures rely on substantially overlapping regions.

\item \textbf{Uniqueness.} For a signature $k$, we define
\begin{equation}
U_K(k) =
1 - \frac{\left| \mathcal{T}_K(k) \cap
\left( \bigcup_{l \neq k} \mathcal{T}_K(l) \right) \right|}
{\left| \mathcal{T}_K(k) \right|}.
\end{equation}
$U_K(k)$ represents the fraction of the top-$K$ patches attributed to signature $k$ that are not shared with any other signature. A high $U_K(k)$ therefore indicates that the regions supporting signature $k$ are spatially distinct from those supporting all the other signatures.
\end{itemize}
Together, these two statistics provide complementary views of spatial specificity: $J_K(k,l)$ assesses separation between pairs of signatures, whereas $U_K(k)$ assesses whether the regions supporting a given signature are distinct from those supporting all other signatures collectively. 

For each tumor type and signature of interest, patients were stratified using
the ground-truth genomic exposures: the positive group comprised patients in
whom the signature was among the top three by exposure, the negative group
those in whom it was not. Because $c_i^{(k)}$ reflects the evidence the model
uses to support its own prediction, both groups were further restricted to
correctly classified samples, so that the attribution maps analyzed correspond
to cases in which the model correctly recovered the presence or absence of the
signature.
The top-$K$ analyses were repeated for $K \in \{10, 20, 50, 100\}$ to assess
the robustness of the results to the number of highly attributed patches
considered. For each tumor type and signature, the distributions of Jaccard
overlap and uniqueness were compared between correctly
classified positive and negative samples using one-sided Mann--Whitney U tests.
The expected direction was lower Jaccard overlap
and higher uniqueness in positive samples, consistent with greater
signature-specific spatial evidence when the mutational process is genomically
active.

\section*{Data availability}
The histopathological whole-slide images and matched whole-genome sequencing data analysed in this study are available from the NCI Genomic Data Commons (GDC) portal (\url{https://portal.gdc.cancer.gov/}). TCGA controlled-access WGS data were obtained under approval of the GDC Data Access Committee; the external cohort was obtained from the CPTAC project through the same portal and approval. 

\section*{Code availability}
The code implementing Hist2Sig, together with trained model weights and scripts to reproduce the main analyses, is available at \url{https://github.com/flavio141/Hist2Sig-paper.git}

\section*{Acknowledgements}
The results shown here are based upon data generated by the TCGA Research Network \url{https://www.cancer.gov/tcga} and by the National Cancer Institute Clinical Proteomic Tumor Analysis Consortium \url{https://proteomic.datacommons.cancer.gov/pdc/}

\section*{Author contributions}
C.P and F.S designed the study. F.S. and C.P. conducted the experiments and performed the analyses. C.P. and F.S. wrote the first draft of the manuscript. P.F., C.R., T.S and I.C. contributed to revising and finalizing the manuscript. P.F and C.P supervised the work. All authors read and approved the final manuscript.

\section*{Competing interests}
The authors declare no competing interests.

\bibliography{sn-bibliography}

\end{document}


\begin{center}
{\LARGE\textbf{Supplementary Material}}\\[1em]
{\Large {Predicting Mutational Signature Exposures from H\&E Whole Slide Images: \\ [1em]
A Pan-Cancer Feasibility Study}}\\[2em]
{\large
Flavio Sartori$^{1}$,
Cesare Rollo$^{2}$,
Isabella Caranzano$^{1}$,
Tiziana Sanavia$^{1}$,  
Piero Fariselli$^{1,\dagger}$,
Corrado Pancotti$^{3,\dagger}$
}\\[1em]

\begin{minipage}{0.85\textwidth}
\centering
\small

$^{1}$Department of Medical Sciences, University of Torino, Via Santena 19, 10123 Torino, Italy\\

$^{2}$Center for Health Data Science, Department of Computer Science, University of Copenhagen, Denmark\\[0.5em]

$^{3}$Helmholtz AI, Helmholtz Munich, Ingolstädter Landstraße 1, 85764 Munich, Germany\\[0.5em]
$^{\dagger}$These authors jointly supervised this work.

\end{minipage}
\end{center}

\vspace{2em}
\newpage

\section*{Supplementary Figures}

\begin{figure}[h!]    \includegraphics[width=1.1\linewidth, clip]{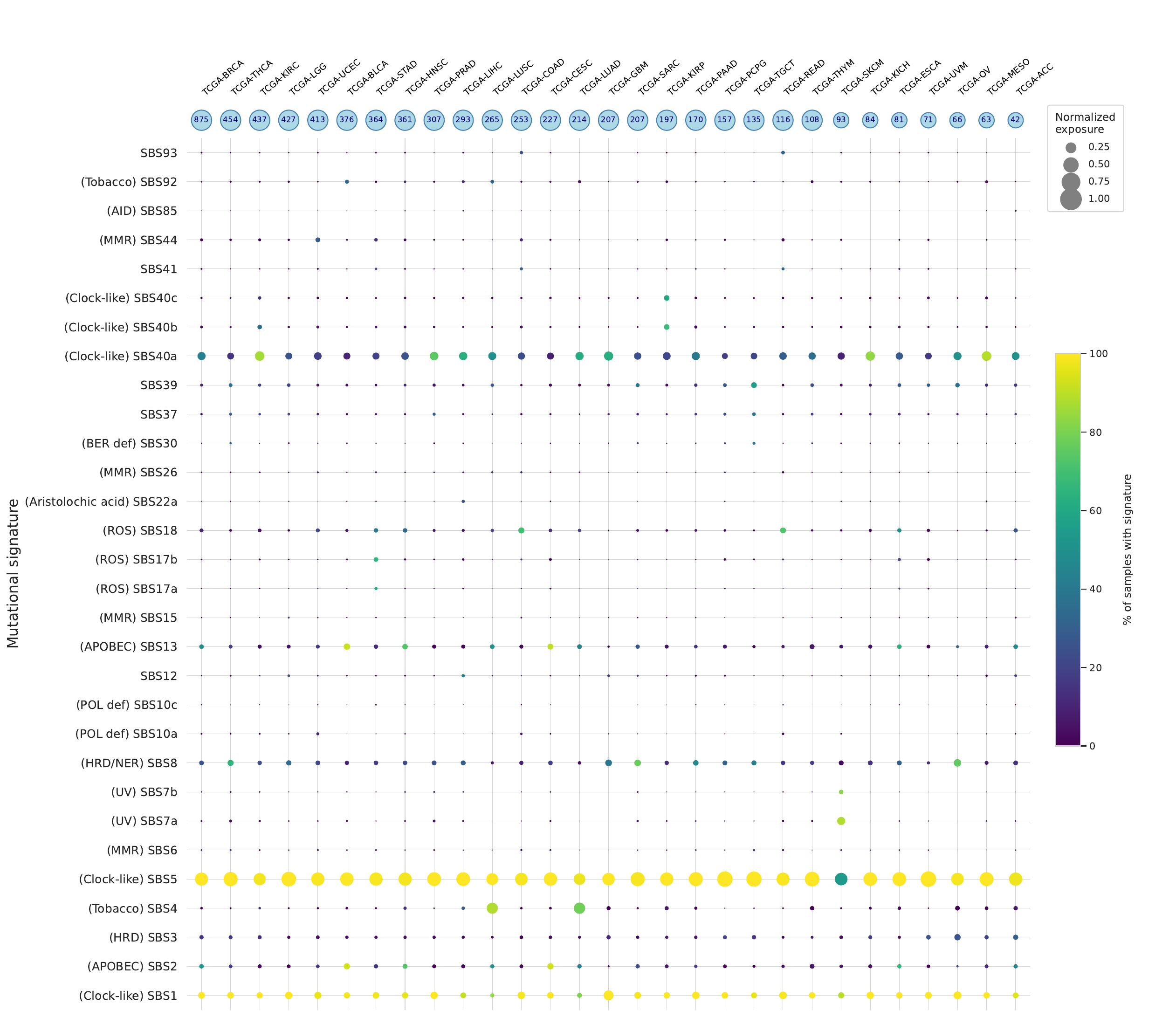}
     \caption{\textbf{Normalized mutational signature exposures across TCGA cancer types}. Each row represents a COSMIC mutational signature, annotated with its etiological process where established, and each column a tumor type ordered by decreasing sample size. Bubble size reflects the mean normalized exposure, and bubble color the percentage of samples with that signature (0–100\%)).}
    \vspace{15pt}  
    \label{fig:TCGA_sig_supp}
\end{figure}

\begin{figure}[h!]    \includegraphics[width=0.8\linewidth, clip]{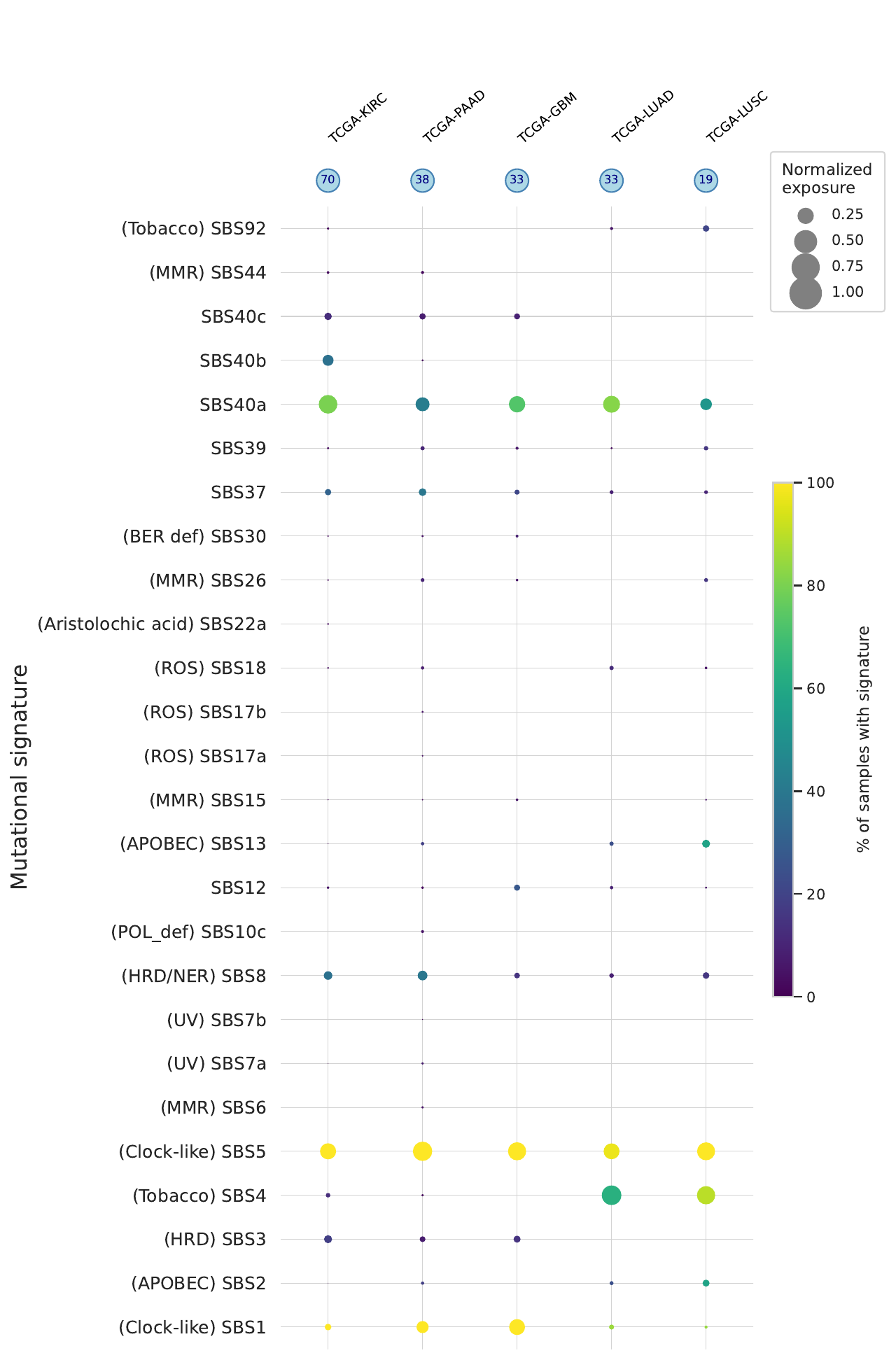}
\caption{\textbf{Normalized mutational signature exposures across CPTAC cancer types}. Each row represents a COSMIC mutational signature, annotated with its etiological process where established, and each column a tumor type ordered by decreasing sample size. Bubble size reflects the mean normalized exposure, and bubble color the percentage of samples with that signature (0–100\%)).}
    \vspace{15pt}  
    \label{fig:CPTAC_sig_supp}
\end{figure}

\begin{figure}[h!]
    \centering
    \includegraphics[width=0.75\linewidth, clip]{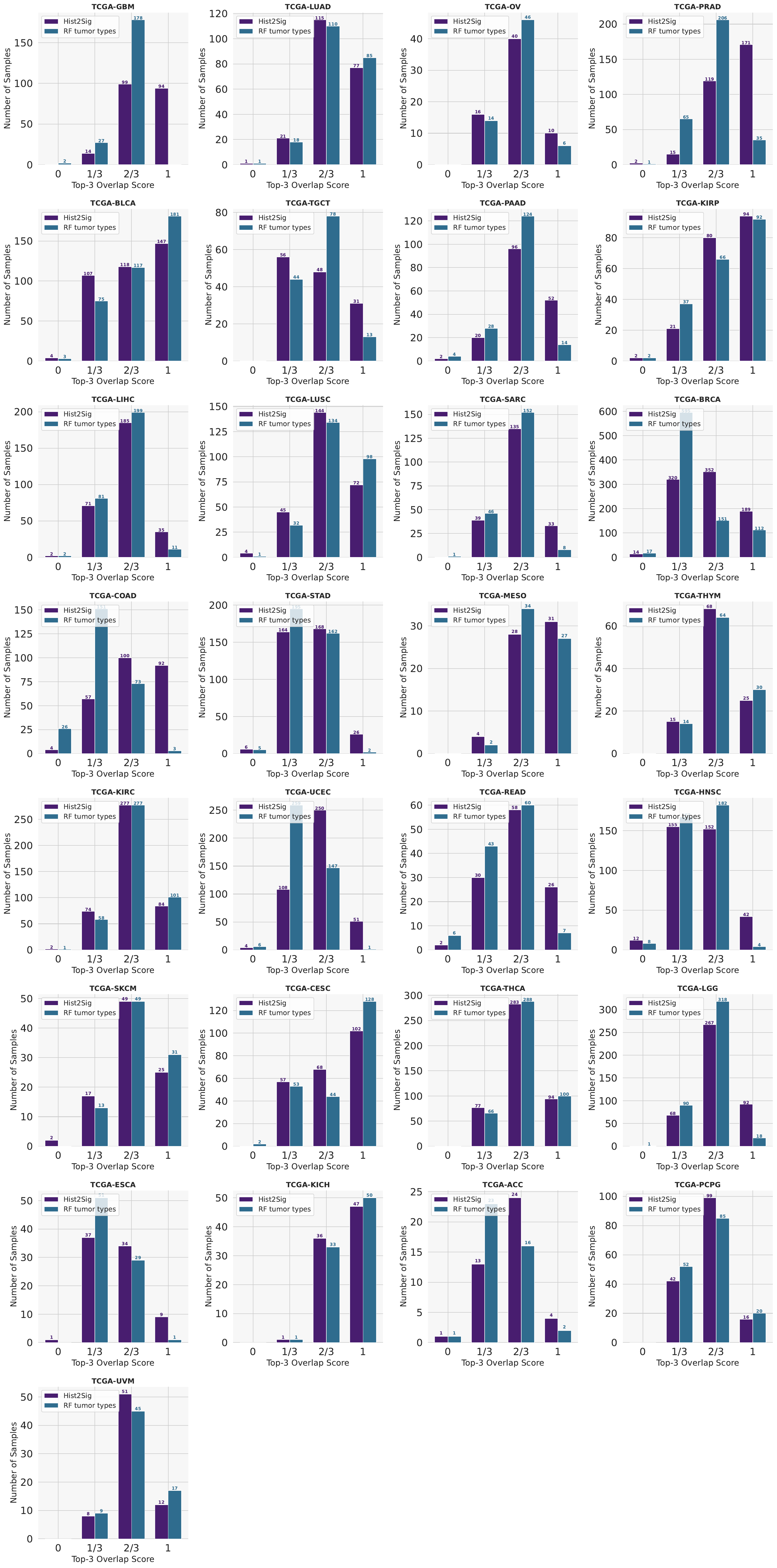}
    \caption{\footnotesize  \textbf{Per sample Top3 overlap score distributions across
    all TCGA tumor types.} Each panel shows the distribution of top 3
    overlap scores for Hist2Sig (violet) and the Random Forest baseline
    trained on tumor type labels (blue). The score represents the
    fraction of the three dominant observed signatures correctly
    identified among the top three predicted, and takes values of 0,
    1/3, 2/3, or 1. Tumor types are ordered by decreasing sample size.}
    \vspace{15pt}  
    \label{fig:topk_complete}
\end{figure}

\begin{figure}[h!]              
    \includegraphics[width=1\linewidth, clip]{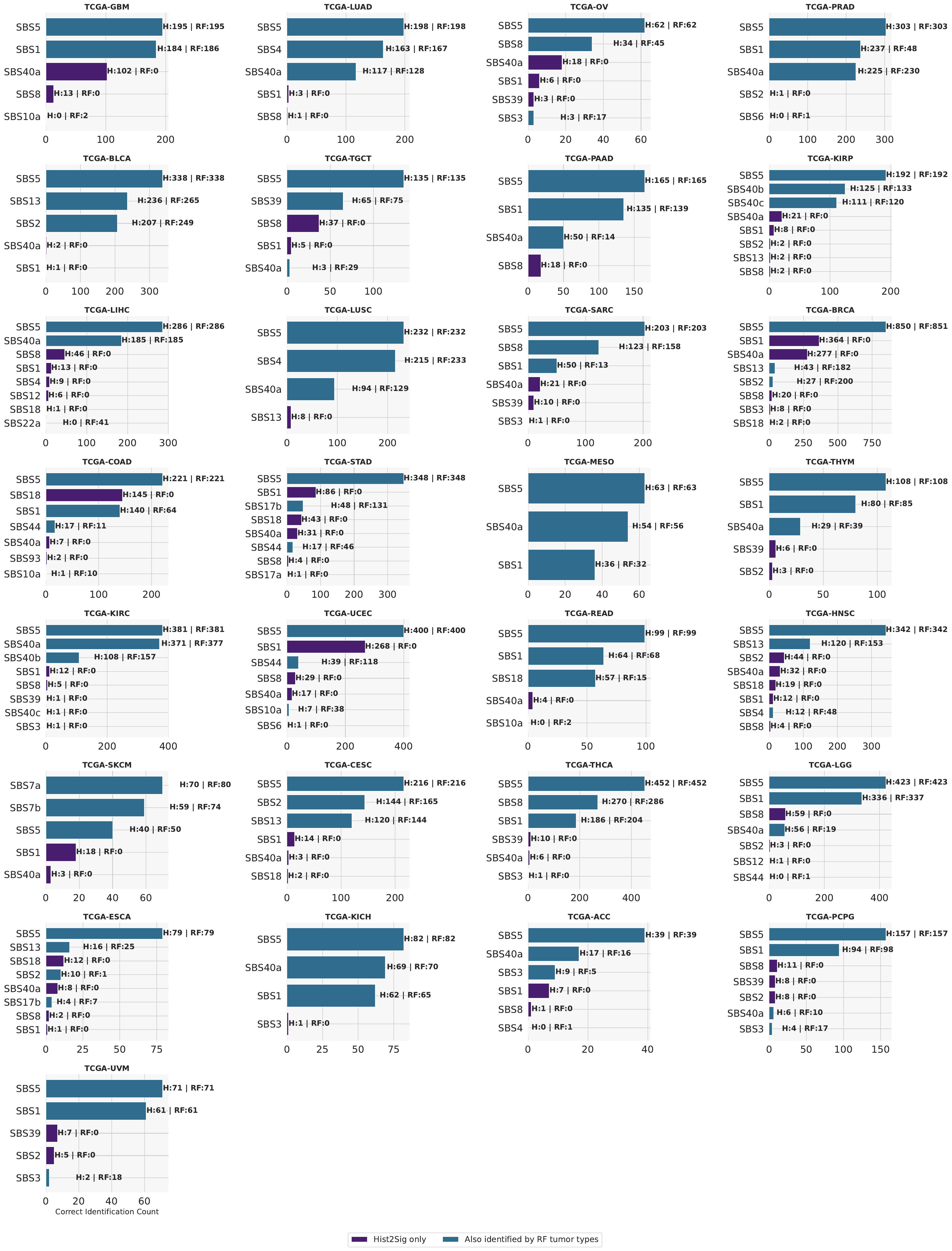}
    \caption{\footnotesize  \textbf{Signature diversity in top-3 predictions across all
    TCGA tumor types.} For each tumor type, horizontal bars show the
    number of samples in which each signature was correctly identified
    among the top three. Violet bars denote signatures recovered
    exclusively by Hist2Sig, Blue bars denote signatures also identified
    by the Random Forest baseline. Counts for each model are reported
    alongside each bar (H: Hist2Sig, RF: Random Forest).}
    \vspace{15pt}  
    \label{fig:diversity_tcga_complete}
\end{figure}

\begin{figure}[h!]    \includegraphics[width=1.\linewidth]{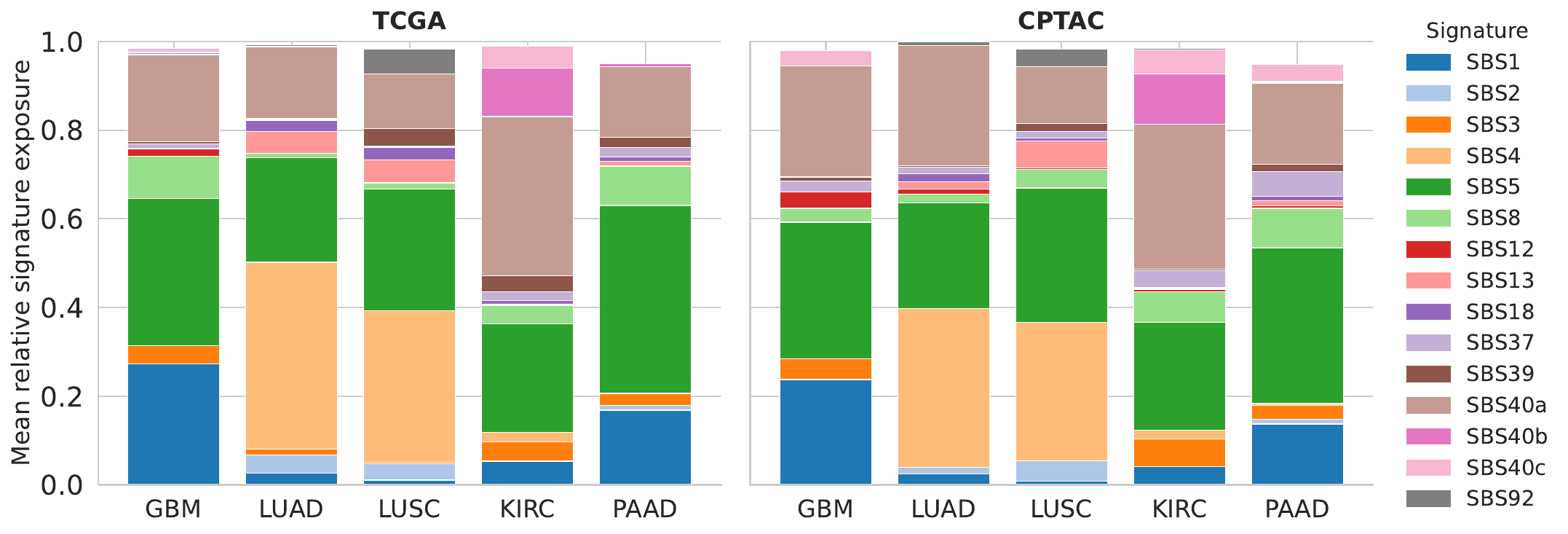}
    \caption{\footnotesize  \textbf{Mean relative signature composition across matched
    tumor types in TCGA and CPTAC.} Each bar shows the mean relative
    exposure to each signature across samples within a tumor type,
    normalized to sum to one. Only signatures contributing more than
    2\% in at least one cohort are shown. The two cohorts display
    broadly similar profiles, supporting the use of CPTAC as a
    biologically matched external validation set.}
    \vspace{15pt}  
    \label{fig:sig_comp_tcga_cpact}
\end{figure}

\begin{figure}[h!]    \includegraphics[width=1.\linewidth]{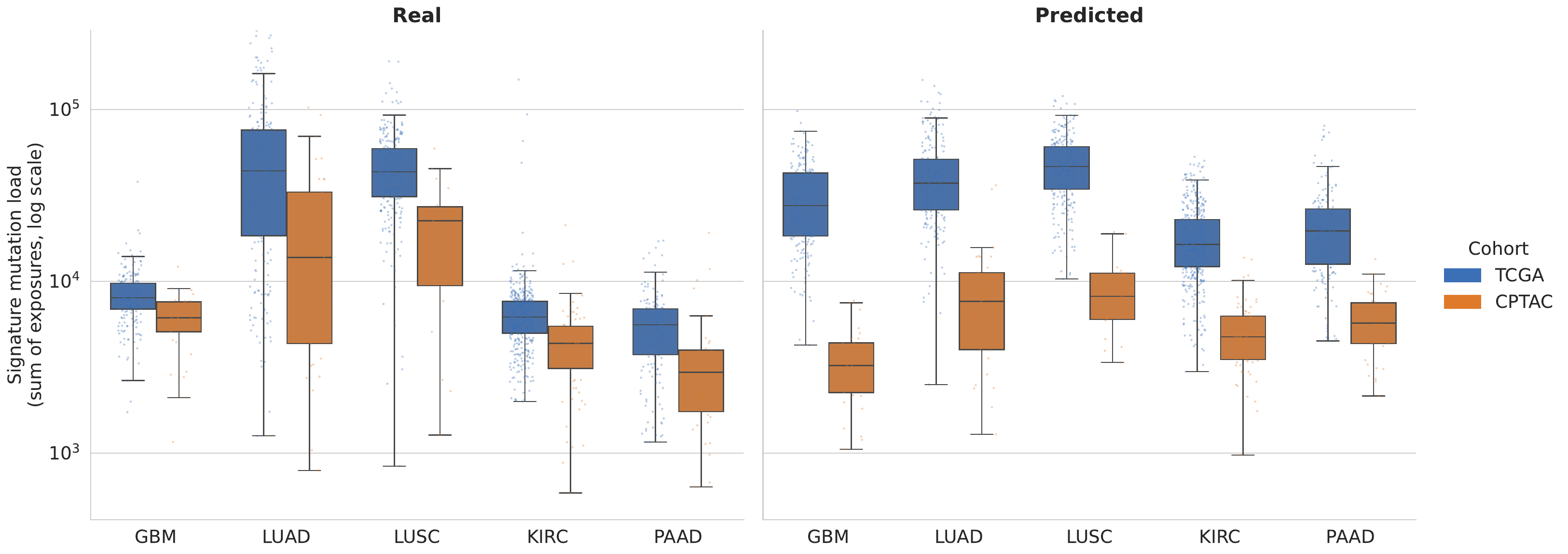}
 \caption{\footnotesize \textbf{Signature mutation load.} Boxplots show the total mutation load, defined as the sum
    of all signature exposure counts, for each of the five tumor types
    shared between the TCGA (blue) and CPTAC (orange) cohorts. The left panel
    displays the observed values derived from SigProfilerAssignment on
    matched whole-genome sequencing data; the right panel displays the
    corresponding Hist2Sig predictions from histopathological images.
    Both panels share the same logarithmic y-axis to allow direct
    comparison. The model preserves the relative ranking of tumor types
    by mutation load and recovers the systematic TCGA--CPTAC gap, but
    amplifies the between-cohort difference, with overestimation in
    TCGA (most visible in GBM) and compression toward lower values in
    CPTAC.}
    \vspace{15pt}  
    \label{fig:tmb_real_pred}
\end{figure}

\clearpage

\begin{figure}[ht!]
\centering
\begin{subfigure}[t]{0.44\linewidth}
    \includegraphics[width=\linewidth]{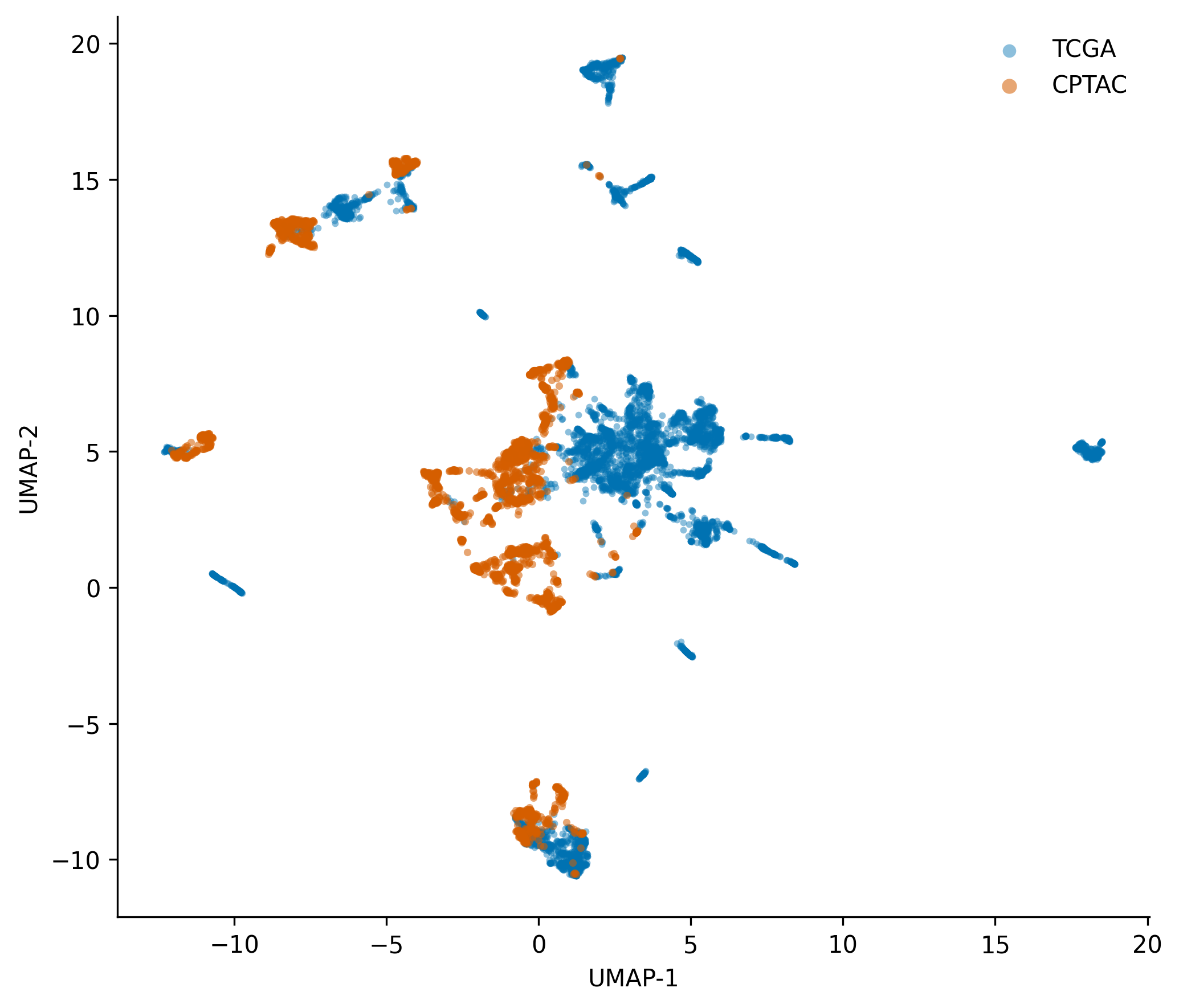}
    \caption{}
    \label{fig:domainshift_dataset}
\end{subfigure}
\hfill
\begin{subfigure}[t]{0.54\linewidth}
    \includegraphics[width=\linewidth]{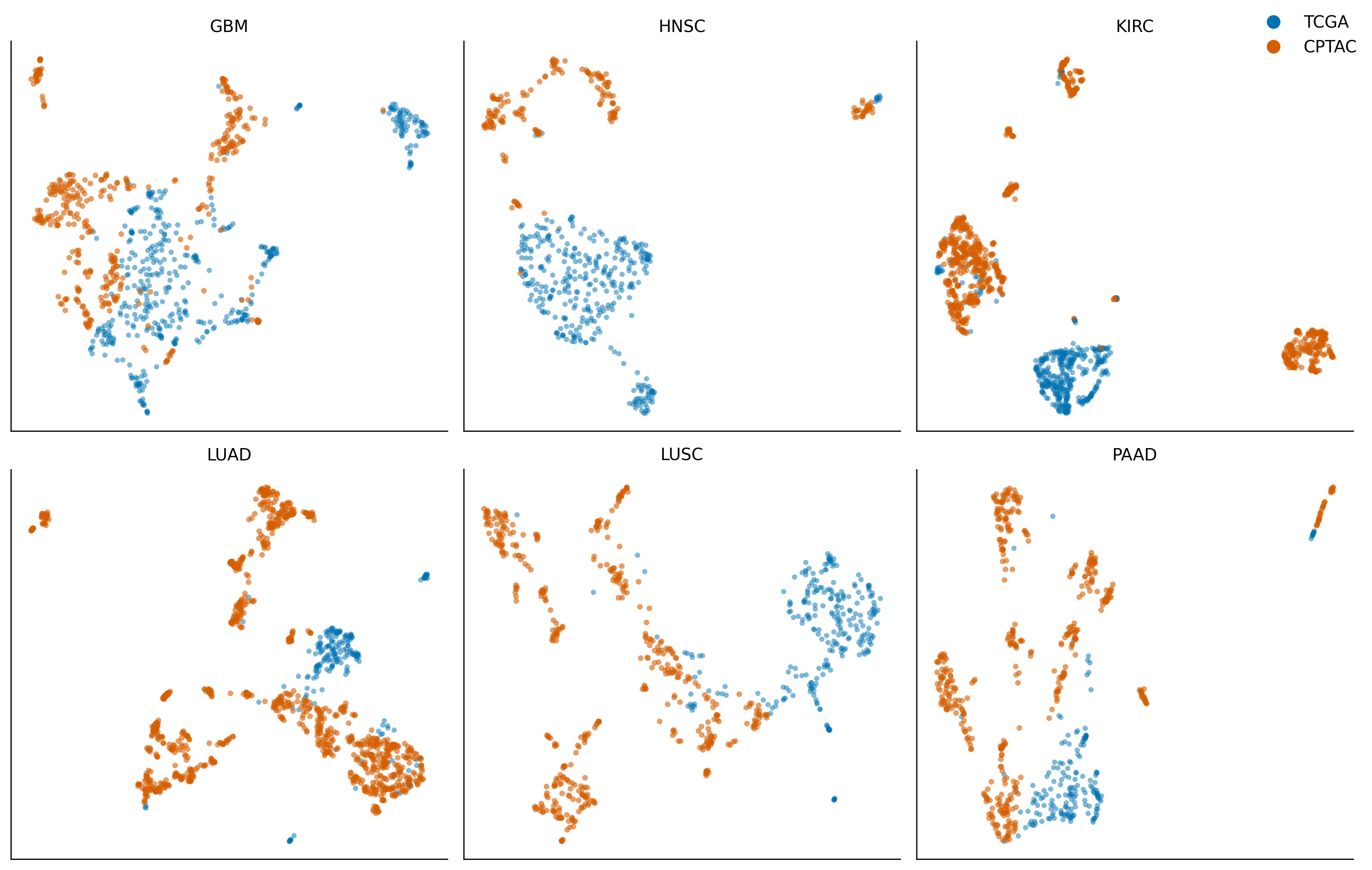}
    \caption{}
    \label{fig:domainshift_per_tumor}
\end{subfigure}
\caption{\footnotesize \textbf{Domain shift between the TCGA and CPTAC cohorts in the
    H-optimus-v1 representation space.}
    Slide-level embeddings were obtained by averaging patch-level features
    within each slide and projected using UMAP with a cosine metric,
    \texttt{n\_neighbors} $=30$, and \texttt{min\_dist} $=0.1$.
    \textbf{(A)} Joint UMAP projection of all included slides, colored by
    cohort of origin. TCGA (blue) and CPTAC (orange) slides occupy largely
    distinct regions of the representation space, indicating a substantial
    global cohort effect.
    \textbf{(B)} Independent UMAP projections computed within each primary
    tumor type represented in both cohorts. Cohort-related displacement is
    visible within GBM, HNSC, KIRC, LUAD, LUSC, and PAAD, although its
    magnitude and structure vary across tumor types. This indicates that
    the observed domain shift cannot be explained solely by differences in
    the distribution of tissues of origin. Because each tumor-specific
    UMAP was fitted independently, the coordinates and relative positions
    are not directly comparable across panels.}
\vspace{15pt}
\label{fig:domainshift}
\end{figure}